\documentclass[5p]{elsarticle}

\makeatletter
\newtoks\@eadauthorshort
\def\@author#1#2{\g@addto@macro\elsauthors{\normalsize%
		\def\baselinestretch{1}%
		\upshape\authorsep#1\unskip\textsuperscript{%
			\ifx\@fnmark\@empty\else\unskip\sep\@fnmark\let\sep=,\fi
			\ifx\@corref\@empty\else\unskip\sep\@corref\let\sep=,\fi
		}%
		\def\authorsep{\unskip,\space}%
		\global\let\@fnmark\@empty
		\global\let\sep\@empty}%
	\@eadauthor={#1}
	\@eadauthorshort={#2}
}
\def\@@author[#1]#2#3{\g@addto@macro\elsauthors{%
		\def\baselinestretch{1}%
		\authorsep#2\unskip\textsuperscript{%#1%
			\@for\@@affmark:=#1\do{%
				\edef\affnum{\@ifundefined{X@\@@affmark}{1}{\elsRef{\@@affmark}}}%
				\unskip\sep\affnum\let\sep=,}%
			\ifx\@fnmark\@empty\else\unskip\sep\@fnmark\let\sep=,\fi
			\ifx\@corref\@empty\else\unskip\sep\@corref\let\sep=,\fi
		}%
		\def\authorsep{\unskip,\space}%
		\global\let\sep\@empty\global\let\@corref\@empty
		\global\let\@fnmark\@empty}%
	\@eadauthor={#2}%
	\@eadauthorshort={#3}%
}
\gdef\@ead#1{\bgroup\def\_{\string\underscorechar\space}%
	\def\{{\string\lbracechar\space}%
	\def~{\hashchar\space}%
	\def\}{\string\rbracechar\space}%
	\edef\tmpA{\the\@eadauthor}
	\edef\tmpB{\the\@eadauthorshort}
	\immediate\write\@auxout{\string\emailauthor
		{#1}{\expandafter\strip@prefix\meaning\tmpA}{\expandafter\strip@prefix\meaning\tmpB}}%
	\egroup
}
\gdef\emailauthor#1#2#3{\stepcounter{ead}%
	\g@addto@macro\@elseads{\raggedright%
		\let\corref\@gobble
		\eadsep\texttt{#1} (\ifemailshortauthor #3\else#2\fi)\def\eadsep{\unskip,\space}}%
}
\newif\ifemailshortauthor
\makeatother

\emailshortauthortrue

\journal{Automatica}

\usepackage{hyperref}
\hypersetup{
	colorlinks   = true, %Colours links instead of ugly boxes
	urlcolor     = blue, %Colour for external hyperlinks
	linkcolor    = blue, %Colour of internal links
	citecolor   = red %Colour of citations
}

\usepackage{mathtools}
\usepackage{amsmath,amssymb,latexsym,amsfonts}
\usepackage{graphicx}

\usepackage{subfigure}
\usepackage{multirow}
\usepackage{enumerate}
\usepackage{calc,cases}
\usepackage{color,soul}
\usepackage[utf8]{inputenc}
\usepackage{gensymb}

\usepackage{color}

\definecolor{red}{rgb}{1,0,0}

\renewcommand{\figurename}{Fig.}

\newcommand{\figref}[1]{\figurename~\ref{#1}}
\newcommand{\tableref}[1]{Table~\ref{#1}}

\newcommand{\secref}[1]{Section~\ref{#1}}

\newcommand{\Realset}{\mathbb{R}}
\newcommand{\z}{s}
\newcommand{\G}{G(\z)}
\newcommand{\C}{C(\z)}

\newcommand{\ov}{\omega_{\nu}}
\newcommand{\Mv}{M_{\nu}}

\usepackage{ifpdf}
\ifpdf
\usepackage{epstopdf}
\DeclareGraphicsExtensions{.pdf,.eps,.png,.jpg,.mps,.bmp}
\else
\usepackage[dvips]{epsfig}
\fi

\biboptions{authoryear}

\begin{document}

\begin{frontmatter}

%\title{A Proportional-multi-resonant tuning based on the Generalized Forced Oscillation method}
%\title{A Ziegler-Nichols forced oscillation method for proportional-multi-resonant control}
%\title{A Generalized Forced Oscillation Method for Tuning Proportional-Multi-resonant Controllers}
\title{A model-free tuning method for proportional-multi-resonant controllers}
\tnotetext[mytitlenote]{This work was supported in part by the Conselho Nacional de Desenvolvimento Científico e Tecnológico (CNPq) and in part by the Coordenação de Aperfeiçoamento de Pessoal de Nível Superior (CAPES), Brazil, Finance Code 001. The material in this paper was not presented at any conference. This paper was recommended for publication in revised form by Associate Editor}

\author[]{Charles~Lorenzini\corref{mycorrespondingauthor}}{C. Lorenzini}
\cortext[mycorrespondingauthor]{Corresponding author}
\ead{lorenzinicharles@gmail.com}

\author[]{Luís~Fernando~Alves~Pereira}{L.~F.~A.~Pereira}
\ead{lfpereira@ufrgs.br}

\author[]{Alexandre~Sanfelice~Bazanella}{A.~S.~Bazanella}
\ead{bazanella@ufrgs.br}

\author[]{Gustavo~R.~Gonçalves~da~Silva}{G.~R.~Gonçalves~da~Silva}
\ead{gustavo.rodrigues.gs@gmail.com}

\address{Graduate Program in Electrical Engineering, Federal University of Rio Grande do Sul, Porto Alegre 90035-190, Brazil}




\begin{abstract}
	 Resonant controllers are widely used in applications involving reference tracking and disturbance rejection of periodic signals. The controller design is typically performed by a trial-and-error approach or by means of time and resource-consuming analytic methods that require an accurate plant model, intricated mathematics and sophisticated tools. In this paper, we propose an easily implementable, model-free method for tuning a proportional-multi-resonant controller applicable to general linear time-invariant causal plants. Just like the Ziegler-Nichols methods, the proposed methodology consist in identifying one specific point of the plant’s frequency
	 response %two parameters of the plant dynamics 
	 -- which is easily obtained in a relay with adjustable phase experiment -- and then designing the controller %from this identified point and the resonance frequencies
with simple tuning formulas and tables. The method is analyzed in detail for three examples, showing its practical appeal and wide applicability.

	 %Driven by applications where the control system should track and/or reject a periodic signal, but no plant model is available to tune the controller, a tuning methodology for proportional-resonant controllers has been recently proposed based on a generalization of the Ziegler-Nichols forced oscillation method which also includes plants that have no ultimate point. Following this same concept, in which only one most appropriate point of the frequency response is identified in a relay with adjustable phase experiment, and considering the case where inherent harmonic frequencies of a periodic signal appear in the loop (whether as a composition of the reference, or as a disturbance), this paper proposes a tuning methodology for a proportional-multi-resonant controller. The proposed method includes the identification experiment and tuning formulas and tables -- just like the Ziegler-Nichols methods -- which provide appropriate stability margin for each class of plants, and is analyzed in detail for three examples, showing its wide applicability.
\end{abstract}


\end{frontmatter}

%\linenumbers

%%%%%%%%%%%%%%%%%%%%%%%%%%%%%%%%%%%%%%%%%%%%%%%%%%%%%%%%%%%%%%%%%%%%%%%%%%%
%%%%%%%%%%%%%%%%%%%%%%%%%%%%%%%%%%%%%%%%%%%%%%%%%%%%%%%%%%%%%%%%%%%%%%%%%%%
%%%%%%%%%%%%%%%%%%%%%%%%%%%%%%%%%%%%%%%%%%%%%%%%%%%%%%%%%%%%%%%%%%%%%%%%%%%
%%%%%%%%%%%%%%%%%%%%%%%%%%%%%%%%%%%%%%%%%%%%%%%%%%%%%%%%%%%%%%%%%%%%%%%%%%%
%%%%%%%%%%%%%%%%%%%%%%%%%%%%%%%%%%%%%%%%%%%%%%%%%%%%%%%%%%%%%%%%%%%%%%%%%%%
%%%%%%%%%%%%%%%%%%%%%%%%%%%%%%%%%%%%%%%%%%%%%%%%%%%%%%%%%%%%%%%%%%%%%%%%%%%

\section{Introduction}

On one hand, resonant controllers are widely employed for applications involving reference tracking and disturbance rejection of periodic signals. Some applications of the resonant controllers involve DC-AC inverters, as in voltage-sources converters %(VSCs)
 \citep{art:res:Teodorescu:2006:ups,art:Castilla:2009:PMR_xi,art:Yepes:2011:Nyquist,art:res:pereira:2014:mr,art:Xin:2018,art:Hans:2020:Design_PMR} and active power filters \citep{art:Lascu:2007:Active:Power:Filters,art:Trinh:2013:APF_Res}. Other applications include vibration control in flexible structures \citep{art:res:Moheimani:2005:ress_struc}, high-precision positioning systems \citep{art:res:Habibullah:2017:vibracao,art:Tao:2020:Nano:PMR}.

On the other hand, %even though resonant controllers are widely used and studied by the scientific community,
there is still no easily understandable, nor easily computable, model-free tuning method that is applicable to this control structure. The controller parameters are often designed by trial-and-error approaches or by means of time and resource-consuming analytic methods that require an accurate plant model, intricated mathematics and sophisticated tools, as can be seen in the references cited above.

Model-free tuning methods for proportional-integral-derivative (PID) controllers, which are appropriate for tracking/rejection of constant signals, were proposed in the seminal work \cite{art:pid:ZN:1942}. These methods consist in identifying two parameters of the plant dynamics, which are easily obtained experimentally, and then designing the controller from these identified parameters from simple tuning formulas. For this reason, the Ziegler-Nichols methods have had a huge influence on control systems design \citep{book:pid:astrom:1995:pid}. 

Recently, there have been developments of model-free tuning methods for resonant controllers. A forced oscillation method for a resonant structure applicable to plants that have an ultimate frequency (that is, whose Nyquist plot crosses the negative real axis) has been proposed in \cite{art:pereira:2015:PR-ZN}. This method can be implemented through the relay feedback experiment \citep{art:astrom:1984:rele} that is generally applied to PID controllers. Then, the parameters of the resonant structure can be computed from simple tuning formulas, similar to the Ziegler-Nichols methods.

This idea has been further extended in our previous work \cite{art:lorenzini:2019:GFO_PR}, where we proposed the Generalized Forced Oscillation (GFO) method for tuning a proportional-resonant (PR) controller. The GFO method can be applied to more general linear time-invariant causal (LTIC) plants, regardless of the existence of an ultimate frequency. This methodology consists in performing the relay with adjustable phase (RAP) experiment, which has been introduced in \cite{art:bazanella:2017:PID-rele-foi} and developed in \cite{art:lorenzini:2019:GFO_PID}, to identify the most appropriate point of the frequency response for each class of plants, and then tuning the PR parameters from this point and the resonance frequency by simple formulas. 

In this paper, we propose a significant development to the GFO method for tuning proportional-multi-resonant (PMR) controllers. The main contribution is the definition of easily computable tuning formulas and tables for the PMR controller \textit{with up to five resonant modes} considering the obtainment of appropriate stability margins and closed-loop performance. %For each resonant mode three parameters are determined, totaling up the design of fifteen parameters.
%-- where three parameters are determined for each resonant mode, that is, totaling up to fifteen parameters designed -- considering the obtainment of appropriate stability margins and closed-loop performance.
Moreover, an analysis of the GFO method with the proposed formulas is performed in three different plants, which indicates their applicability to a wide variety of plants with different characteristics. Hence, we show that the GFO method is a sound and convenient way to tune PMR controllers without the need of a plant model and with little design effort, using only one
simple experiment and previously obtained tuning formulas and tables. %, which can be implemented by a layman of control or in auto-tuning applications.

%This paper is organized as follows. Preliminary concepts are shown in \secref{sec:prel}, where we define the control problem, the design philosophy,  and also the class of plants we are dealing with. The design problem formulation
%is described in \secref{sec:problem}, which introduces the PMR controller and devises the limitations and the constraints involved in the GFO method to tune this controller. In \secref{sec:variables}, we determine the tuning variables and also the batch of experiments that yielded the tuning tables proposed in \secref{sec:tables}. The applicability of the GFO method with the proposed tuning tables is verified in \secref{sec:bench_plants} through an %detailed 
%analysis with three different plants.

%%%%%%%%%%%%%%%%%%%%%%%%%%%%%%%%%%%%%%%%%%%%%%%%%%%%%%%%%%%%%%%%%%%%%%%%%%%
%%%%%%%%%%%%%%%%%%%%%%%%%%%%%%%%%%%%%%%%%%%%%%%%%%%%%%%%%%%%%%%%%%%%%%%%%%%
%%%%%%%%%%%%%%%%%%%%%%%%%%%%%%%%%%%%%%%%%%%%%%%%%%%%%%%%%%%%%%%%%%%%%%%%%%%
%%%%%%%%%%%%%%%%%%%%%%%%%%%%%%%%%%%%%%%%%%%%%%%%%%%%%%%%%%%%%%%%%%%%%%%%%%%
%%%%%%%%%%%%%%%%%%%%%%%%%%%%%%%%%%%%%%%%%%%%%%%%%%%%%%%%%%%%%%%%%%%%%%%%%%%
%%%%%%%%%%%%%%%%%%%%%%%%%%%%%%%%%%%%%%%%%%%%%%%%%%%%%%%%%%%%%%%%%%%%%%%%%%%

\section{Preliminaries} \label{sec:prel}

%%%%%%%%%%%%%%%%%%%%%%%%%%%%%%%%%%%%%%%%%%%%%%%%%%%%%%%%%%%%%%%%%%%%%%%%%%%
%%%%%%%%%%%%%%%%%%%%%%%%%%%%%%%%%%%%%%%%%%%%%%%%%%%%%%%%%%%%%%%%%%%%%%%%%%%

\subsection{Control problem} \label{subsec:contr_probl}

We consider LTIC plants in a closed-loop feedback control system, which can be represented by
\begin{equation}
\label{eq:plant}
\begin{aligned}
Y(\z) {} = {} & \G U(\z), \\
E(\z) {}={} R(\z) - Y(\z),&~~~U(\z) {}={} \C E(\z),
\end{aligned}
\end{equation}
where $\G$ is the transfer function of a strictly proper either bounded-input, bounded-output (BIBO)-stable or type $1$ plant\footnote{We consider type 1 plants the ones possessing one pole at the origin.}. $U(\z)$, $Y(\z)$, $R(\z)$, and $E(\z)$ are respectively the Laplace transforms of the control input, the plant's output -- the controlled variable --, the reference, and the tracking error. $\C$ is the controller transfer function.% to be determined. 

The main objective is to derive tuning formulas for the parameters of a given structure of $\C$ considering tracking/rejection of composite periodic signals, without knowing the plant model, but only one specific point of its frequency response. Tuning methods that require obtainment of only one point of the plant's frequency response are usually based on forced oscillation experiments.

%\subsection{Controller} \label{subsec:contr}
%, i.e., $G(j\ov)=\Mv e^{j\nu}$. 

%The controller tuning requirement consists in moving this point to an specific point $p$ in the complex plane, that is, the loop function satisfies
%\begin{equation}
%G(j\ov)C(j\ov)=p=M_{\rho}e^{j\rho}.
%\label{eq:loop_spec}
%\end{equation}
%
%Next, we discuss the specification of $\ov$ and the point $G(j\ov)$ for different classes of plants, followed by the definition of the controller structure. Then, in Section \ref{sec:variables}, we define the tuning variable $p$ for each class of plants and for a given amount $N$ of resonant controllers.

%%%%%%%%%%%%%%%%%%%%%%%%%%%%%%%%%%%%%%%%%%%%%%%%%%%%%%%%%%%%%%%%%%%%%%%%%%%
%%%%%%%%%%%%%%%%%%%%%%%%%%%%%%%%%%%%%%%%%%%%%%%%%%%%%%%%%%%%%%%%%%%%%%%%%%%

\subsection{Tuning methods based on forced oscillation} \label{sec:CFO}

Controller tuning based on forced oscillation is a well-known procedure since Ziegler-Nichols' tuning methods and formulas for PID controllers were proposed in \cite{art:pid:ZN:1942}. Essentially, this procedure consists in identifying the plant's ultimate point, i.e., the point at which its Nyquist plot crosses the negative real axis -- corresponding to the lowest frequency where its phase is $-180\degree$ -- and then designing the controller parameters to move this point to a predefined place in the complex plane. The relay feedback experiment \citep{art:astrom:1984:rele} -- that under certain conditions yields a sustained oscillation at the plant's output -- is the most common way to identify the plant's ultimate point. We will refer to the combination of the relay feedback experiment with the tuning formulas proposed in \cite{art:pid:ZN:1942} as the Classical Forced Oscillation (CFO) method. %Over the years, different locations $p$ have been proposed in order to obtain different performance and stability margins; the tuning formulas presented in \cite{art:pid:ZN:1942} correspond to $p = -0.4+j0.08$ and $p = -0.6 -j0.28$, respectively for PI and PID controllers. 
An overview of the CFO method and some of its extensions are presented in \cite{book:pid:astrom:1995:pid}.

However, a large number of plants does not posses an ultimate point -- all minimum-phase stable first and second-order plants, and most plants with relative degree smaller than three -- and so are not amenable to the application of the CFO method. Moreover, these methods were classically limited to PID tuning. In this case, one can still design the controller to more generic LTIC plants based on other relevant frequency response points, which can be identified through a relay with adjustable phase (RAP) experiment \citep{art:bazanella:2017:PID-rele-foi,art:lorenzini:2019:GFO_PID}; we thus called it the Generalized Forced Oscillation (GFO) method. The case of a PID structure using the GFO method is presented in \cite{art:lorenzini:2019:GFO_PID} and for the PR structure in \cite{art:lorenzini:2019:GFO_PR}. %, and for the PMR in \blu{\cite{art:lorenzini:2020:GFO_PMR}}.
%A PID tuning method that is applicable to a large class of plants that have no ultimate point and uses a relay with adjustable phase RAP experiment was presented in \cite{art:bazanella:2017:PID-rele-foi} and further developed in \cite{art:lorenzini:2019:GFO_PID}. A CFO-based methodology for a resonant structure has been proposed in \cite{art:pereira:2015:PR-ZN}, considering plants that have an ultimate point, and this idea has been further developed in \cite{art:lorenzini:2019:GFO_PR} for a PR controller that can be applied to more generic LTIC plants.

Based on the theoretical approach of the CFO method, the GFO method consists in identifying the point of the plant's frequency response at which the phase reaches a previously specified value $\nu$:
\begin{equation}
\label{eq:Gwv_res}
G(j\omega_{\nu}) = M_\nu \angle \nu = M_\nu \left(\cos\left(\nu\right) + j\sin\left(\nu\right)\right),
\end{equation}
that is, determine the quantities $\omega_\nu$ and $M_\nu$:
\begin{equation}
\label{eq:Gwv}
\omega_\nu = \underset{\omega \geq 0}{\textnormal{min}}\; \omega : \angle G(j\omega) = \nu \;\;\; \textnormal{and}\;\;\;M_\nu = |G(j\omega_\nu)|.
\end{equation}

Once these quantities are somehow obtained, then design the controller parameters such that
\begin{equation}
\label{eq:CGwv_p_gene}
C(j\omega_{\nu})G(j\omega_{\nu}) = p = M_\rho \left(\cos\left(\rho\right) + j\sin\left(\rho\right)\right),
\end{equation}	
or equivalently,
\begin{equation}
\label{eq:Cwv_p_gene}
C(j\omega_{\nu}) = \frac{M_\rho}{M_\nu} \left(\cos(\rho-\nu) + j\sin(\rho-\nu)\right),
\end{equation}	
where $p$ is a previously specified location in the complex plane.

As we did in our previous works \cite{art:lorenzini:2019:GFO_PID,art:lorenzini:2019:GFO_PR}, we divide the LTIC plants in three classes, as follows, according to the identified point.

%Following our previous works \cite{art:lorenzini:2019:GFO_PID,art:lorenzini:2019:GFO_PR}, we present next a subdivision of the plants in three classes, according to the identified point.

%In what follows, PR tuning formulas will be proposed through the solution of \eqref{eq:Cwv_p_gene} for the transfer function \eqref{eq:C_pr_xi} with $s = j\omega_{\nu}$. We restrict our presentation to the case where $\omega_r<\omega_\nu$.

%Tracking a reference at such a high frequency does not make much practical sense, so considering also the case 
%$\omega_r>\omega_\nu$ would complicate 

%%%%%%%%%%%%%%%%%%%%%%%%%%%%%%%%%%%%%%%%%%%%%%%%%%%%%%%%%%%%%%%%%%%%%%%%%%%
%%%%%%%%%%%%%%%%%%%%%%%%%%%%%%%%%%%%%%%%%%%%%%%%%%%%%%%%%%%%%%%%%%%%%%%%%%%

\subsubsection{Classes of plants} \label{subsec:class_plants}

For plants with an ultimate point, this is clearly  the point that must be used, so the choice $\nu=-180^o$ is self-evident for this class, which for future convenience will be called Class A. Thus, in Class A, the ultimate point of the plant's frequency response is identified, that is, 
\begin{equation}
\label{eq:Gw_180}
\begin{aligned}
&\nu = -180\degree,\; \omega_\nu = \omega_u = \underset{\omega \geq 0}{\textnormal{min}}\,\omega : \angle G(j\omega) = -180\degree,\\
&M_\nu = M_u  = |G(j\omega_u)| = 1/K_u. 
\end{aligned}
\end{equation}
%As for the choice of the location to which the ultimate point should be shifted, parameter $p$, we took  
%the classical ZN point for PI tuning ($p=-0.4+\jmath 0.08$) as a first approximation. After numerous tests, we have found the best results in terms of stability margins and closed-loop performance with:
%\begin{equation}
%\label{eq:p_a_m3}
%p = 0.4\left(\cos(-183\degree)+j \sin(-183\degree)\right), 
%\end{equation}
%for $0<\omega_r/\omega_u<0.5$, and
%\begin{equation}
%\label{eq:p_a_m1}
%p = 0.4\left(\cos(-181\degree)+j \sin(-181\degree)\right) 
%\end{equation}
%for $0.5 \leq \omega_r/\omega_u < 1$.

%If the plant has no ultimate point and considering that $\omega_r<\omega_\nu$, then the gain margin will be infinite, provided that the  controller zeros are located in adequate positions such that the controller does not contribute with a too large phase lag. Thus, the phase margin becomes a control design objective for plants that have no ultimate point \cite{art:bazanella:2017:PID-rele-foi}. Based on the theoretical approach of the CFO method, in order to tune controllers applied to plants that have no ultimate point the GFO's idea  is to identify a specific point of the plant's frequency response, and then design a controller to achieve a desired phase margin at this particular frequency.

Consider now the plants that do not possess an ultimate point. In this case we devise two frequency points based on the plant phase response.
%In previous work \cite{art:bazanella:2017:PID-rele-foi} it was found that $\nu=-120^o$
%was the best choice for the tuning of PID controllers for plants without an ultimate point. %As for $p$, it has been proposed in  \cite{art:bazanella:2017:PID-rele-foi}  to pick it such that a phase margin of $50^o$ was achieved, which corresponds to $p = 1 \angle -130^o$. We have successfully tested these same choices here for the tuning of PR controllers, so this is what  we propose, provided that the plant's frequency response achieves this phase for some frequency. 
The first set of plants is given when $\nu=-120\degree$, that is, those that do not possess an ultimate point but whose frequency response reaches $-120\degree$ for some frequency. This set will herein be called Class B.
Thus, for plants in Class B we have 
\begin{equation}
\label{eq:Gw_120}
\begin{aligned}
&\nu = -120\degree,\,
\omega_\nu = \omega_{120} = \underset{\omega \geq 0}{\textnormal{min}}\, \omega \!:\! \angle G(j\omega) = -120\degree,\\
&M_\nu = M_{120} = |G(j\omega_{120})|.
\end{aligned}
\end{equation}

Finally, consider those plants whose frequency response never reaches $-120\degree$, which will be
called Class C in this paper. The Class C plants is such that:
\begin{equation}
\label{eq:Gw_60}
\begin{aligned}
&\nu = -60\degree,\,
\omega_\nu = \omega_{60} = \underset{\omega \geq 0}{\textnormal{min}}\, \omega \!:\! \angle G(j\omega) = -60\degree,\\
&M_\nu = M_{60} = |G(j\omega_{60})|.
\end{aligned}
\end{equation}
%\begin{equation}
%\label{eq:p_pr_c}
%p = 1 \angle -90\degree = \cos(-90\degree) + j\sin(-90\degree),
%\end{equation}
%which provide a phase margin of $90\degree$, assuming that the loop transfer function is monotonically decreasing for frequencies higher than $\omega_{60}$.

In practical situations where the plant model is unknown, these specific frequency points can be identified using the RAP experiment, as in \cite{art:lorenzini:2019:GFO_PID,art:lorenzini:2019:GFO_PR}, which is briefly described next.

%%%%%%%%%%%%%%%%%%%%%%%%%%%%%%%%%%%%%%%%%%%%%%%%%%%%%%%%%%%%%%%%%%%%%%%%%%%
%%%%%%%%%%%%%%%%%%%%%%%%%%%%%%%%%%%%%%%%%%%%%%%%%%%%%%%%%%%%%%%%%%%%%%%%%%%

\subsubsection{Relay with adjustable phase (RAP) experiment}\label{sec:rap}

\begin{figure}[t!]
	\centering
	\includegraphics[width=0.70\columnwidth]{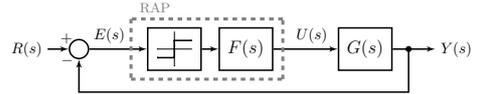}
	\caption{Block diagram of the RAP experiment.}
	\label{fig:RAP}
\end{figure}

Consider the experiment setup presented in \figref{fig:RAP}, which consists of a relay and where $F(s)$ is a known transfer function with \textit{constant phase} $\gamma$ in a defined range of frequencies, i.e., $\angle F(j\omega) = \gamma, \forall \,\omega\in\omega_{min}<\omega<\omega_{max}$, as proposed in \cite{art:bazanella:2017:PID-rele-foi}. To obtain a transfer function with flat phase frequency response $\gamma$ that is not necessarily an entire multiple of $-90\degree$, an approximation of a fractional order integrator (FOI) \citep{book:tepljakov:2017:fractional} is employed. The FOI has transfer function $\hat{F}(s) = 1/(s^m)$, whose phase is given by $\angle \hat{F}(j\omega) = -90\degree \times m$. Thus, for a given $m = -\gamma/90\degree$, $\angle \hat{F}(j\omega) = \gamma$ is obtained, as desired.

If the relay phase is chosen such that $\gamma  \triangleq - 180\degree -\nu$ and if the relay gain is adjusted so that the self-oscillation condition is achieved, one obtains the ultimate frequency $\ov$ of the transfer function $F(s)G(s)$, i.e., $\ov \!:\!\angle F(j\ov)G(j\ov) = -180\degree$. Thus, the \textit{plant's magnitude and phase at this frequency} can be approximately computed as:
\begin{center}
	$\left|G(j\ov)\right| {} = {} \dfrac{\pi A_\nu}{4 d \left|F(j\ov)\right|},\, \angle G(j\ov) {} = {} \nu {} = {} -180\degree - \gamma$,
\end{center}
where $d\in\Realset^+$ is the relay gain, $A_\nu$ is the oscillation amplitude at the plant's output and $\left|F(j\ov)\right|$ is the FOI magnitude at $\omega_\nu$ (see \cite{art:lorenzini:2019:GFO_PID,art:lorenzini:2019:GFO_PR} for details).

To identify the points of the plant's frequency response defined in Subsection~\ref{subsec:class_plants}, the RAP experiment is started with $\gamma = 0\degree$, i.e.,  the traditional relay experiment is performed, so a Class A plant can be identified. If a self-oscillation condition is not obtained %, that is, the plant's output is not a well-defined signal with oscillation's amplitude and period constants over time, 
then variable $\gamma$ must be decreased from $0\degree$ to $-60\degree$, allowing identification of a Class B plant. If still a self-oscillation condition is not obtained, then decrease $\gamma$ from $-60\degree$ to $-120\degree$ and a Class C plant is identified. %Thus, the RAP experiment enables an automatic procedure for identification of a defined point of the plant's frequency response -- whether having or not an ultimate point -- in a single experiment without designer intervention.

Since we have formalized the plants we are dealing with, their subdivisions in three different classes, the tuning philosophy -- the GFO method -- and how to obtain these specific frequency points -- the RAP experiment --, the one thing left to define before the tuning variables and tables is the controller structure that will be tuned using these concepts, and this is where the contributions of this work start. 

%%%%%%%%%%%%%%%%%%%%%%%%%%%%%%%%%%%%%%%%%%%%%%%%%%%%%%%%%%%%%%%%%%%%%%%%%%%
%%%%%%%%%%%%%%%%%%%%%%%%%%%%%%%%%%%%%%%%%%%%%%%%%%%%%%%%%%%%%%%%%%%%%%%%%%%
%%%%%%%%%%%%%%%%%%%%%%%%%%%%%%%%%%%%%%%%%%%%%%%%%%%%%%%%%%%%%%%%%%%%%%%%%%%
%%%%%%%%%%%%%%%%%%%%%%%%%%%%%%%%%%%%%%%%%%%%%%%%%%%%%%%%%%%%%%%%%%%%%%%%%%%
%%%%%%%%%%%%%%%%%%%%%%%%%%%%%%%%%%%%%%%%%%%%%%%%%%%%%%%%%%%%%%%%%%%%%%%%%%%
%%%%%%%%%%%%%%%%%%%%%%%%%%%%%%%%%%%%%%%%%%%%%%%%%%%%%%%%%%%%%%%%%%%%%%%%%%%

\section{Main contributions} \label{sec:problem}

\subsection{Controller} \label{subsec:controller}

Before defining the controller structure used in this work, let us introduce some reasoning behind it. Following the  internal model principle \citep{art:contr:francis:1975}, a stable closed-loop tracks/rejects asymptotically a periodic signal with fundamental frequency $\omega_r$ if the controller $\C$ has poles at $\pm j n \omega_r$, where $n$ represents the order of the harmonics that compose the signal of interest, assuming they are not zeros of $\G$. This characteristic results in controllers with multiple resonance peaks with infinite magnitude at the frequencies $n \omega_r$, hence the denomination of resonant controller.

Besides, it is often the case where the periodic signal of interest is decomposed in an infinite sum of sinusoid signals with frequencies multiple integer of $\omega_r$. Clearly, practical implementation of such a controller is infeasible, and usually a finite number of resonant modes are tuned in the frequencies with larger contribution of the signal of interest \citep{art:res:pereira:2014:mr,art:res:Teodorescu:2006:ups}. Moreover, resonant modes shifted -- by means of a damping factor $\xi_n$ -- to the semicircle of radius $n \omega_r$ in the left-half plane are often used, yielding finite magnitude at the frequencies $n \omega_r$, making the tuning easier and improving robustness \citep{art:Castilla:2009:PMR_xi,art:res:Teodorescu:2006:ups}. On one hand, these two aspects -- finite resonant modes and magnitude -- yield non-null steady-state error due to partially compensated components. On the other hand, when the finite PMR structure is well-tuned, most of performance requirements are attained in typical applications.

Finally, it has been shown in \cite{art:lorenzini:2019:GFO_PR} that, using a single PR controller for Class A processes, when $\omega_r \approx \omega_u$ the controller structure presents very large gains in a range around the plant's ultimate frequency and stability margins are much smaller than when $\omega_r \ll \omega_u$, so a small change in the controller contribution to the phase at this specific frequency significantly changes the stability margins and the closed-loop performance. Because of that, in that work two sets of tuning variables were defined for this specific case, depending on the ratio $\omega_r/\omega_u$. To overcome the need to devise two tuning points and provide better stability margins for this class of plants, here we consider also a phase-lead block in the controller, that is to be used for Class A processes.

Thus, in this work we consider the following transfer function for the PMR controller:
\begin{equation}
\C = \underbrace{\left(k_a\frac{s+z_a}{s+p_a}\right)}_{C_{l}(s)}\prod_{n}\underbrace{\left(K_{p_n}+\frac{K_{r_{1n}}s+K_{r_{2n}}}{s^2+2\xi_n\omega_{rn}s+\omega_{rn}^2}\right)}_{C_{pr_n}(s)},
\label{eq:cpmr}
\end{equation}
where $\{k_a,z_a,p_a\}$ compose a phase-lead block $C_{l}(s)$ that will be only adjusted for Class A plants, $\omega_{rn}=n\omega_r$ are the frequencies composing the periodic signal to be tracked and/or rejected, $\xi_n$ are the damping coefficients of the resonant modes, $\{K_{p_n},K_{r_{1n}},K_{r_{2n}}\}$ are the parameters to be tuned for each $n$ in $C_{pr_n}(s)$.

Notice that instead of the PMR controller parallel configuration, as in \cite{art:Castilla:2009:PMR_xi,art:Xin:2018},
we have opted for the series configuration in \eqref{eq:cpmr}, given by the product of PR structures at multiple-integer frequencies of $\omega_r$. This configuration is advantageous in the development of its tuning formulas, since each PR structure can be designed independently for a given tuning point.

Equating \eqref{eq:Cwv_p_gene} to $C_{pr_{n}}(s)$ in \eqref{eq:cpmr}, for $s=j\omega_\nu$, and solving for the variables $K_{p_n}$, $K_{r_{1n}}$ and $K_{r_{2n}}$, one can verify a degree of freedom for the PR controller parameters, since there are two equations to be solved, one for the real part and another for the imaginary part. Following the idea of our previous work \cite{art:lorenzini:2019:GFO_PR}, we introduce a third equation involving a constraint for the controller zeros of each intermediate PR structure, such that their product satisfies $\eta_n^2\omega_{r}^2$, where $0\!<\!\eta_n\!\leq\!1$ is a parameter to be determined next. This constraint yields at least one zero with module less than $\omega_r$ -- when they are real --, or a pair of complex conjugate zeros with module $\eta_n \omega_r$.

Thus, the parameters of the PR structure regarding the harmonic of order $n$ are determined by the following set of generalized tuning formulas:
\begin{equation}
\label{eq:kpn_gain}
\begin{aligned}
K_{p_n}{}={} &\frac{M_{\rho_n}(\ov^2-n^2\omega_r^2)\cos(\rho_n-\nu)}{\Mv(\ov^2-\eta_n^2n^2\omega_r^2)} \\ 
&+{}\frac{2\ov\xi_n n\omega_r\sin(\rho_n-\nu)}{\Mv(\ov^2-\eta_n^2n^2\omega_r^2)}, \\
K_{r_{1n}}= &-\frac{M_{\rho_n}(\ov^2-n^2\omega_r^2)\sin(\rho_n-\nu)}{\Mv\ov} \\ 
&- \frac{M_{\rho_n} 2\xi_n n^3\omega_r^3(\eta_n^2-1)\cos(\rho_n-\nu)}{\Mv(\ov^2-\eta_n^2n^2\omega_r^2)} \\
&-\frac{4\ov\xi_n^2 n^2\omega_r^2\sin(\rho_n-\nu)}{\Mv(\ov^2-\eta_n^2n^2\omega_r^2)}, \\
K_{r_{2n}}=&\frac{M_{\rho_n}(\ov^2 - n^2\omega_r^2)\cos(\rho_n-\nu)(\eta_n^2-1)n^2\omega_r^2}{\Mv(\ov^2 - \eta_n^2n^2\omega_r^2)} \\
& + \frac{M_{\rho_n}2\ov\xi_n n^3\omega_r^3\sin(\rho_n-\nu)(\eta_n^2-1)}{\Mv(\ov^2 - \eta_n^2n^2\omega_r^2)} \cdot
\end{aligned}
\end{equation}

The set of formulas \eqref{eq:kpn_gain} compose a general formulation for the PMR structure \eqref{eq:cpmr} for any given amount -- herein denoted by $N$ -- of resonant modes of order $n$. Let us now define the limitations of the analysis and synthesis of the PMR parameters provided in this work, and some constraints to be applied in order to ease the formulation.

%%%%%%%%%%%%%%%%%%%%%%%%%%%%%%%%%%%%%%%%%%%%%%%%%%%%%%%%%%%%%%%%%%%%%%%%%%%
\subsection{Limitations} \label{ssec:limitations}

The synthesis of the tuning formulas for the PMR controller will be developed considering up to five,  i.e., $N=5$, resonant modes. For most practical applications that require resonant controllers, this is enough \cite{art:res:pereira:2014:mr,art:res:Habibullah:2017:vibracao,art:Xin:2018}. We will also consider the following two sets of resonant modes on the analysis:
\begin{enumerate}[(i)]
	\item resonant modes in $\omega_r,~2\omega_r,~3\omega_r,~4\omega_r,~5\omega_r$; 
	\item resonant modes in $\omega_r,~3\omega_r,~5\omega_r,~7\omega_r,~9\omega_r$.
\end{enumerate}
This is because the Continuous-time Fourier Series representation of a periodic signal shows that it can be written as sum of sinusoid terms in multiple integers of the fundamental frequency. The first set appears, for example, when decomposing a sawtooth wave signal, whereas the second set appears in decomposition of square and triangle wave signals, and in disturbances applied to voltage-sources converters. %voltage-source converters. %\hl{se a UPS for citada em algum momento na intro, dai coloca aqui tbm.}

Moreover, this set of tuning formulas will be developed for the generic case $\xi_n\geq 0$ and will be restricted to $\max(n\omega_r)<\ov$. 

\subsection{Constraints} \label{ssec:constraints}

Since the whole PMR structure is a product of intermediate PR ones, the location $p$ in the complex plane for the tuning equation \eqref{eq:CGwv_p_gene} is decomposed in intermediate locations. For example, for the PMR controller with $3$ resonant modes, one has
\begin{equation}
\text{(i)} ~~p = p_1\times p_2\times p_3, ~~~~\text{or}~~~~ \text{(ii)} ~~p = p_1\times p_3\times p_5.
\label{eq:p_intermediate}
\end{equation}

This decomposition in intermediate locations must verify the following module and phase conditions:
\begin{align}
\label{eq:mod_cond} M_\rho&=\prod_n M_n, \\
\label{eq:phase_cond} \rho&=\sum_n \rho_n.
\end{align}
When considering also the phase-lead block, its contribution $p_{l}=M_{l}e^{j\rho_{l}}$ should be included in these constraints. 

%\hl{na tese eh o paragrafo que comeca com: assim, a equacao geral de sintonia do metodo GOF.} {\color{red} Não entendi.}, mas aqui tem que explicar um pouco mais, e citar daí o outro artigo.

The magnitude adjustment with the PMR controllers will be made in the first resonant mode -- this choice simplifies and facilitates the development of the tuning formulas. Thus, taking into account the condition \eqref{eq:mod_cond}  in \eqref{eq:Cwv_p_gene} we have
\begin{equation}
|C_{pr_1}(j\ov)|=\frac{M_{\rho_1}}{\Mv},
\label{eq:mag_c1}
\end{equation}
whereas for the remaining $C_{pr_n}(s)$ and $C_{l}(s)$, it follows
\begin{eqnarray*} 
	|C_{pr_n}(j\ov)|=|C_{l}(j\ov)| = 1, ~n>1;\\
	\Mv = M_{\rho_n} = M_{l}=1, ~n>1.
\end{eqnarray*}

Therefore, it remains to specify, for each $C_{pr_n}(s)$, the phase contribution at frequency $\ov$ and the relative position of its zeros, that is, the parameters $\rho_n$ and $\eta_n$, plus the value of $\rho_l$. Thus, in this paper the development of the tuning formulas \eqref{eq:kpn_gain} will consider: 
\begin{itemize}
	\item the three different classes of plants - A, B and C defined in Section II;
	\item the PMR structure (\ref{eq:cpmr}) for up to 5 modes;
	\item the two sets of resonant modes - cases (i) and (ii) defined above.
\end{itemize}

%%%%%%%%%%%%%%%%%%%%%%%%%%%%%%%%%%%%%%%%%%%%%%%%%%%%%%%%%%%%%%%%%%%%%%%%%%%
%%%%%%%%%%%%%%%%%%%%%%%%%%%%%%%%%%%%%%%%%%%%%%%%%%%%%%%%%%%%%%%%%%%%%%%%%%%
%%%%%%%%%%%%%%%%%%%%%%%%%%%%%%%%%%%%%%%%%%%%%%%%%%%%%%%%%%%%%%%%%%%%%%%%%%%
%%%%%%%%%%%%%%%%%%%%%%%%%%%%%%%%%%%%%%%%%%%%%%%%%%%%%%%%%%%%%%%%%%%%%%%%%%%
%%%%%%%%%%%%%%%%%%%%%%%%%%%%%%%%%%%%%%%%%%%%%%%%%%%%%%%%%%%%%%%%%%%%%%%%%%%
%%%%%%%%%%%%%%%%%%%%%%%%%%%%%%%%%%%%%%%%%%%%%%%%%%%%%%%%%%%%%%%%%%%%%%%%%%%

\subsection{Tuning variables definition} \label{sec:variables}

In order to carefully define the variables used for adjusting \eqref{eq:cpmr}, the design of the PMR controller was evaluated for $18$ families of LTIC processes belonging to classes A, B and C. These processes are the same used in \cite{art:lorenzini:2019:GFO_PR} and taken from \cite{art:bazanella:2017:PID-rele-foi,book:pid:astrom:1995:pid}, summing up $123$ plants for Class A, and $98$ plants for classes B and C. For each process class, different combinations of the variables $\rho_n$ and $\eta_n$ were examined for each control topology with up to five resonant modes in the two frequency sets.

First, let us start by defining the location $p$ for each process class, taking into account our previous work on a single PR structure \citep{art:lorenzini:2019:GFO_PR}, which considered the batch of experiments described above. For a Class A process, we started from $p = 0.4e^{-j181\degree}$ and after testing different locations and because here we considered the phase-lead block, we propose the following location:
\begin{align} 
p&=p_{l} \times p_{pmr} = e^{j46.4^{\circ}} \times 0.4e^{-j187^{\circ}} =0.4e^{-j140.6^{\circ}} %\\
\label{eq:p_classA}
\end{align}
where $p_{l}$ and $p_{pmr}$ are the tuning points for the $C_{l}(s)$ and $C_{pr}(s)$ controllers, respectively. This location results in gain margin higher than $2.5$ and phase margin of approximately $45\degree$ for $\omega_r \approx \omega_u$.
%\red{I'd remove the following sentence. Also, can't you include a sentence justifying the choice of p, as the
%jury asked to do in the thesis?}
%The insertion of a phase-lead block for Class A process allowed the definition of only one single location $p$, contrary to our previous work \cite{art:lorenzini:2019:GFO_PR}, where two locations were defined.

For a Class B process, we have
\begin{equation}
p = e^{-j130^{\circ}}.
\label{eq:p_classB}
\end{equation}
And finally for a Class C process:
\begin{equation}
p = e^{-j90^{\circ}}. 
\label{eq:p_classC}
\end{equation}
The location $p$ for classes B and C were the same defined in \cite{art:lorenzini:2019:GFO_PR}, which results in phase margin of respectively $50\degree$ and $90\degree$ for each class of processes.

The analysis range of the variable $\rho_n$ for each PR structure is limited by the phase contribution at $\omega_{\nu}$, that is,
$$-\pi/2<\rho_n<0.$$ 
Moreover, it should verify \eqref{eq:phase_cond} -- including $\rho_{l}$ for Class A process -- for the phase of the locations $p$ defined in \eqref{eq:p_classA}, \eqref{eq:p_classB} and \eqref{eq:p_classC} for their respective class. Besides, by definition, $\eta_n$ is limited by $0<\eta_n\leq1$.  

Taking into account the set of generic tuning formulas in \eqref{eq:kpn_gain}, the constraints in Subsection~\ref{ssec:constraints}, the limitations on $\rho_n$ and $\eta_n$, the three classes of plants and their tuning locations \eqref{eq:p_classA}, \eqref{eq:p_classB} and \eqref{eq:p_classC}, the batch of tests in a wide array of plants, and also maximum overshoot of $15\%$ with the periodic references defined in \secref{sec:bench_plants}, we now propose a set of tuning variables for each order $n$ PR controller that constitutes the PMR structure. 

Tables~\ref{tab:var_pr}~and~\ref{tab:var_pmr} present the parameters that yielded better results considering both stability margins and transient performance. From these variable definitions, it is clear that two sets of tuning formulas must be derived: one exclusively for $C_{pr_1}(s)$ as a function of the amount $N$ of resonant modes and for each class, and one single set for the multiple resonant modes, i.e., for $n>1$. In the next section, we propose a general tuning equation for each parameter of the controller \eqref{eq:cpmr} and use the sets of variables presented in tables \ref{tab:var_pr} and \ref{tab:var_pmr} to develop two sets of tuning tables for the PMR controller, considering each process class.

\begin{table}[t]
	\centering \renewcommand{\tabcolsep}{1.9pt}
	\caption{Variables used in \eqref{eq:kpn_gain} for the PR structures}
	\small
	\begin{tabular}{ccccc}
		\hline
		\textbf{\begin{tabular}[c]{@{}c@{}}Harmonic order\end{tabular}} & \textbf{Parameter} & \textbf{Class A} & \textbf{Class B} & \textbf{Class C} \\ \hline
		& $\nu$              & $-180^{\circ}$   & $-120^{\circ}$   & $-60^{\circ}$    \\
		$n=1$                                                                    & $\Mv$          & $M_u$            & $M_{120}$        & $M_{60}$         \\
		& $\ov$              & $\omega_u$       & $\omega_{120}$   & $\omega_{60}$    \\ \hline
		& $\nu$              & $0^{\circ}$      & $0^{\circ}$      & $0^{\circ}$      \\
		$n>1$                                                                    & $\Mv$          & $1$              & $1$              & $1$              \\
		& $\ov$              & $\omega_u$       & $\omega_{120}$   & $\omega_{60}$    \\ \hline
	\end{tabular}
	\label{tab:var_pr}
\end{table}

\begin{table}[t]
	\centering \renewcommand{\tabcolsep}{1.5pt}
	\caption{Tuning variables for the PMR controller}
	\small
	\begin{tabular}{ccc} \hline
		\textbf{Class A} & \textbf{Class B} & \textbf{Class C} \\ \hline
		$p_{l}\prod p_n=0.4e^{-j140.6^{\circ}}$ & $\prod p_n=e^{-j130^{\circ}}$ & $\prod p_n=e^{-j90^{\circ}}$ \\ \hline
		\begin{tabular}[c]{@{}l@{}}$p_1=0.4e^{-j(188-N)^{\circ}}$\\ $\eta_1=\begin{cases}0.6, N =1 \\ 0.7,2\leq N\leq 3 \\ 0.9, N>3 \end{cases}$\end{tabular} & \begin{tabular}[c]{@{}l@{}}$p_1=e^{-j(131-N)^{\circ}}$\\ $\eta_1=\begin{cases}0.7, N =1 \\ 0.9, N\geq2 \end{cases}$\end{tabular} & \begin{tabular}[c]{@{}l@{}}$p_1=e^{-j(91-N)^{\circ}}$\\ $\eta_1=0.9$\end{tabular} \\ \hline
		\multicolumn{3}{c}{$p_n=e^{-j1^{\circ}}~~\text{and}~~\eta_n=0.9,~n>1$} \\ \hline
		%	\multicolumn{3}{c}{{\color{red}Tudo na mesma coluna ou três colunas com a mesma coisa?}} \\ \hline
	\end{tabular}
	\label{tab:var_pmr}
\end{table}

%%%%%%%%%%%%%%%%%%%%%%%%%%%%%%%%%%%%%%%%%%%%%%%%%%%%%%%%%%%%%%%%%%%%%%%%%%%
%%%%%%%%%%%%%%%%%%%%%%%%%%%%%%%%%%%%%%%%%%%%%%%%%%%%%%%%%%%%%%%%%%%%%%%%%%%
%%%%%%%%%%%%%%%%%%%%%%%%%%%%%%%%%%%%%%%%%%%%%%%%%%%%%%%%%%%%%%%%%%%%%%%%%%%
%%%%%%%%%%%%%%%%%%%%%%%%%%%%%%%%%%%%%%%%%%%%%%%%%%%%%%%%%%%%%%%%%%%%%%%%%%%
%%%%%%%%%%%%%%%%%%%%%%%%%%%%%%%%%%%%%%%%%%%%%%%%%%%%%%%%%%%%%%%%%%%%%%%%%%%
%%%%%%%%%%%%%%%%%%%%%%%%%%%%%%%%%%%%%%%%%%%%%%%%%%%%%%%%%%%%%%%%%%%%%%%%%%%

\subsection{Tuning tables} \label{sec:tables}

In this section we propose the sets of tuning equations of the PMR controller \eqref{eq:cpmr} using the GFO method, for the three classes of processes and considering up to 5 resonant modes. Given the set of generalized tuning formulas \eqref{eq:kpn_gain} and the variables $\nu$, $\Mv$, $\ov$, $p_n$, $\eta_n$ defined in tables \ref{tab:var_pr} and \ref{tab:var_pmr}, we propose two sets of particular tuning equations for the PMR controller. The first set is with respect to the controller $C_{pr_1}(s)$ for a given amount $N$ of resonant modes. The second one is a general set for each resonant controller $C_{pr_n}(s)$ with $n>1$. Besides, we restrict the resonant modes frequencies to the interval $0<\max(n\omega_r)< \ov$.

%In this section we present the sets of tuning formulas of the PMR controller \eqref{eq:cpmr} using the GFO method, for the three classes of processes and considering up to 5 resonant modes. Besides, we restrict the resonant modes frequencies to the interval $0<\max(n\omega_r)< \ov$.

%Given this set of generalized tuning formulas \eqref{eq:kpn_gain} and the variables $\nu,\Mv,\ov,p_n,\eta_n$ defined in tables \ref{tab:var_pr} and \ref{tab:var_pmr}, we now present two sets of particular tuning equations for the PMR controller. The first set is with respect to the controller $C_{pr_1}(s)$ for a given amount $N$ of resonant modes. The second one is a general set for each resonant mode with $n>1$.

\subsubsection{Tuning of controller $C_{l}(s)$} 
First, let us define the phase-lead transfer function that yields the location $\rho_l$ in \eqref{eq:p_classA}, for Class A process only:
\begin{equation}
C_{l}(s)=2.5\;\frac{s+0.4\omega_u}{s+2.5\omega_u},
\label{eq:pl_block}
\end{equation}
whose parameters -- the pole and zero locations and also the gain -- were chosen to provide the maximum lead phase contribution of $46.4\degree$ with unitary magnitude at $\omega_u$.

\subsubsection{Tuning of controller $C_{pr_n}(s)$} 
Consider the following parametrized version of \eqref{eq:kpn_gain}:
%
%\begin{align}
%\label{eq:Cpr1_param}
%K_{p_1}&=\frac{\alpha_1(\ov^2-\omega_r^2)}{\Mv(\ov^2-\alpha_3\omega_r^2)}-\frac{\alpha_2\ov\omega_r\xi_1}{\Mv(\ov^2-\alpha_3\omega_r^2)}, \nonumber\\ 
%K_{r_{11}}&=\frac{\beta_1(\ov^2-\omega_r^2)}{\Mv\ov}+\frac{\beta_2\omega_r^3\xi_1+\beta_3\ov\omega_r^2\xi_1^2}{\Mv(\ov^2-\alpha_3\omega_r^2)}, \nonumber\\ 
%K_{r_{21}}&=\frac{\zeta_1\omega_r^2(\omega_r^2-\ov^2)}{\Mv(\ov^2-\alpha_3\omega_r^2)}-\frac{\zeta_2\ov\omega_r^3\xi_1}{\Mv(\ov^2-\alpha_3\omega_r^2)}\cdot
%\end{align}
\begin{equation}
\label{eq:Cprn_param}
\begin{aligned}
K_{p_n}&=\frac{\alpha_1(\ov^2-n^2\omega_r^2)}{\Mv(\ov^2-\alpha_3n^2\omega_r^2)}-\frac{\alpha_2 n \omega_r \ov\xi_n}{\Mv(\ov^2-\alpha_3n^2\omega_r^2)}, \\ 
K_{r_{1n}}&=\frac{\beta_1(\ov^2-n^2\omega_r^2)}{\Mv\ov}+\frac{\beta_2n^3\omega_r^3\xi_n+\beta_3n^2\omega_r^2\ov\xi_n^2}{\Mv(\ov^2-\alpha_3n^2\omega_r^2)}, \\ 
K_{r_{2n}}&=\frac{\zeta_1n^2\omega_r^2(n^2\omega_r^2-\ov^2)}{\Mv(\ov^2-\alpha_3n^2\omega_r^2)}+\frac{\zeta_2n^3\omega_r^3\ov\xi_n}{\Mv(\ov^2-\alpha_3n^2\omega_r^2)}\cdot
\end{aligned}
\end{equation}

Thus, for a given number of resonant modes used in the design, the parameters $\alpha_1,~\alpha_2,~\alpha_3,~\beta_1,~\beta_2,~\beta_3,~\zeta_1,~\zeta_2$ are defined in tables \ref{tab:param_c1} and \ref{tab:param_cn} for $n=1$ and $n>1$, respectively. In an application, considering a class of plants, the controller $C_{pr_1}(s)$ is tuned with the parameters for $n=1$, which are presented in columns 3 to 5 of \tableref{tab:param_c1}, for a given  $N$, whereas the controllers $C_{pr_n}(s)$ for $n>1$ are designed using columns 3 to 5 of \tableref{tab:param_cn}.

Recall that the damping coefficient $\xi_n$ must be chosen by the designer, considering that a stable closed-loop system asymptotically tracks/rejects a given sinusoidal reference/disturbance with frequency $\omega_{rn}$ if the order $n$ PR controller with $\xi_n = 0$ is inserted in the loop.

\begin{table}[t!]
	\centering \renewcommand{\tabcolsep}{4.3pt}
	\caption{Tuning table for controller $C_{pr_1}(s)$}
	\small
	\begin{tabular}{ccccc} \hline
		\textbf{N}           & \textbf{Param.}   &\textbf{Class A} & \textbf{Class B}  & \textbf{Class C} \\ \hline
		\multirow{2}{*}{1 to 5}              & $\Mv$             & $M_u$           & $M_{120}$         & $M_{60}$         \\
		           & $\ov$             & $\omega_u$      & $\omega_{120}$    & $\omega_{60}$ \\ \hline
		\multirow{8}{*}{1}   & $\alpha_1$        & $0.397$  & $0.985$ & $0.866$   \\
		                     & $\alpha_2$        &$0.0975$ & $0.347$ & $1.00$    \\
		                     & $\alpha_3$        &$0.360$  & $0.490$ & $0.810$   \\
		                     & $\beta_1$         &$0.0487$  & $0.174$  & $0.500$    \\
		                     & $\beta2$          &$0.508$   & $1.00$   & $0.329$    \\
		                     & $\beta_3$         &$0.195$   & $0.695$  & $2.00$     \\
		                     & $\zeta_1$         &$0.254$   & $0.502$  & $0.165$    \\
		                     & $\zeta_2$         &$0.0624$  & $0.177$  & $0.190$    \\ \hline
		\multirow{8}{*}{2}  & $\alpha_1$         & $0.398$  & $0.988$ & $0.875$   \\
		                    & $\alpha_2$         & $0.0836$ & $0.313$ & $0.970$   \\
		                    & $\alpha_3$         & $0.490$  & $0.810$ & $0.810$   \\
		                    & $\beta_1$          & $0.0418$  & $0.156$  & $0.485$    \\
		                    & $\beta2$           & $0.406$   & $0.375$  & $0.332$    \\
		                    & $\beta_3$          & $0.167$   & $0.626$  & $1.94$     \\
		                    & $\zeta_1$          & $0.203$   & $0.188$  & $0.166$    \\
		                    & $\zeta_2$          & $0.0426$  & $0.0594$ & $0.184$    \\ \hline
		\multirow{8}{*}{3}  & $\alpha_1$         & $0.398$  & $0.990$ & $0.883$   \\
	                       & $\alpha_2$            & $0.0697$ & $0.278$ & $0.939$   \\
	                       & $\alpha_3$            & $0.490$  & $0.810$ & $0.810$   \\
	                       & $\beta_1$             & $0.0349$  & $0.139$  & $0.469$    \\
	                       & $\beta2$              & $0.406$   & $0.376$  & $0.336$    \\
	                       & $\beta_3$             & $0.139$   & $0.557$  & $1.88$     \\
	                       & $\zeta_1$             & $0.203$   & $0.188$  & $0.168$    \\
	                       & $\zeta_2$             & $0.0356$  & $0.0529$ & $0.178$    \\ \hline
		\multirow{8}{*}{4} & $\alpha_1$                  & $0.399$  & $0.993$ & $0.891$   \\
	                      & $\alpha_2$                     & $0.0558$ & $0.244$ & $0.908$   \\
	                      & $\alpha_3$                     & $0.810$  & $0.810$ & $0.810$   \\
	                      & $\beta_1$                      & $0.0279$  & $0.122$  & $0.454$    \\
	                      & $\beta2$                       & $0.152$   & $0.377$  & $0.339$    \\
	                      & $\beta_3$                      & $0.112$   & $0.487$  & $1.82$     \\
	                      & $\zeta_1$                      & $0.0758$  & $0.189$  & $0.169$    \\
	                      & $\zeta_2$                      & $0.0106$  & $0.0463$ & $0.173$    \\ \hline
		\multirow{8}{*}{5}    & $\alpha_1$              & $0.399$  & $0.995$ & $0.899$   \\
		                      & $\alpha_2$              & $0.0419$ & $0.209$ & $0.877$   \\
		                      & $\alpha_3$              & $0.810$  & $0.810$ & $0.810$   \\
		                      & $\beta_1$               & $0.0209$  & $0.105$  & $0.438$    \\
		                      & $\beta2$                & $0.152$   & $0.378$  & $0.342$    \\
		                      & $\beta_3$               & $0.0837$  & $0.418$  & $1.75$     \\
		                      & $\zeta_1$               & $0.0759$  & $0.189$  & $0.171$    \\
		                      & $\zeta_2$               & $0.00796$ & $0.0397$ & $0.167$    \\ \hline
	\end{tabular}
	\label{tab:param_c1}
\end{table}

%\hl{recall Mv e $\ov$ for each class. Talk about $\xi_1$}
%
%\subsection{$C_{pr_n}(s)$ tuning formulas for $n>1$}
%The set of tuning formulas of $C_{pr_n}(s)$ for $n>1$ obtained by substituting the parameters of tables \ref{tab:var_pr} and \ref{tab:var_pmr} in \eqref{eq:kpn_gain} is given by:

%\begin{align}
%\label{eq:Cpmrn_class_ABC_xi}
%K_{pn}    {} = {} & \frac{\omega_\nu^2 - n^2 \omega_r^2}{\omega_\nu^2 - 0.81 n^2 \omega_{r}^2} -\frac{0.0349 \omega_\nu n \omega_r \xi_n}{\omega_\nu^2 - 0.81 n^2 \omega_{r}^2}, \nonumber\\
%K_{r_{1n}} {} = {} & \frac{0.0175 \left(\omega_\nu^2 - n^2 \omega_{r}^2\right)}{\omega_\nu} \nonumber\\ &+ \frac{0.380 n^3 \omega_r^3 \xi_n + 0.0698 \omega_\nu n^2 \omega_r^2 \xi_n^2}{\omega_\nu^2 -0.81 n^2 \omega_{r}^2}, \nonumber\\
%K_{r_{2n}} {} = {} &  \frac{0.190 n^2 \omega_{r}^2 \left(n^2 \omega_{r}^2 -\omega_\nu^2\right)}{\omega_\nu^2 - 0.81 n^2 \omega_{r}^2} + \frac{0.00663  \omega_\nu n^3 \omega_r^3  \xi_n}{\omega_\nu^2 - 0.81 n^2 \omega_{r}^2}
%\end{align}

\begin{table}[h!]
	\centering \renewcommand{\tabcolsep}{4.0pt}
	\caption{Tuning table for controller $C_{pr_n}(s)$ with $n>1$} \label{tab:param_cn}
	\small
	\begin{tabular}{ccccc } \hline
	\textbf{Harmonic} & \multirow{2}{*}{\textbf{Param.}} & \multirow{2}{*}{\textbf{Class A}}  & \multirow{2}{*}{\textbf{Class B}}  & \multirow{2}{*}{\textbf{Class C}} \\ 
		\textbf{order}          & &   &   &   \\ \hline
		\multirow{10}{*}{$n>1$} & $\Mv$      & $1.00$            & $1.00$            & $1.00$         \\
		                        & $\ov$      & $\omega_u$        & $\omega_{120}$    & $\omega_{60}$ \\
		                        & $\alpha_1$ & $1.00$            & $1.00$            & $1.00$    \\
		                        & $\alpha_2$ & $0.0349$ & $0.0349$ & $0.0349$  \\
		                        & $\alpha_3$ & $0.810$  & $0.810$  & $0.810$   \\
		                        & $\beta_1$  & $0.0175$  & $0.0175$  & $0.0175$   \\
		                        & $\beta2$   & $0.380$   & $0.380$   & $0.380$    \\
		                        & $\beta_3$  & $0.0698$  & $0.0698$  & $0.0698$   \\
		                        & $\zeta_1$  & $0.190$   & $0.190$   & $0.190$    \\ 
		                        & $\zeta_2$  & $0.00663$ & $0.00663$ & $0.00663$  \\ \hline
	\end{tabular}
	
\end{table}

In the following, the GFO method for tuning PMR controllers is applied to three different process. %, each one from a different class.

%%%%%%%%%%%%%%%%%%%%%%%%%%%%%%%%%%%%%%%%%%%%%%%%%%%%%%%%%%%%%%%%%%%%%%%%%%%
%%%%%%%%%%%%%%%%%%%%%%%%%%%%%%%%%%%%%%%%%%%%%%%%%%%%%%%%%%%%%%%%%%%%%%%%%%%
%%%%%%%%%%%%%%%%%%%%%%%%%%%%%%%%%%%%%%%%%%%%%%%%%%%%%%%%%%%%%%%%%%%%%%%%%%%
%%%%%%%%%%%%%%%%%%%%%%%%%%%%%%%%%%%%%%%%%%%%%%%%%%%%%%%%%%%%%%%%%%%%%%%%%%%
%%%%%%%%%%%%%%%%%%%%%%%%%%%%%%%%%%%%%%%%%%%%%%%%%%%%%%%%%%%%%%%%%%%%%%%%%%%
%%%%%%%%%%%%%%%%%%%%%%%%%%%%%%%%%%%%%%%%%%%%%%%%%%%%%%%%%%%%%%%%%%%%%%%%%%%

\section{Illustrative examples} \label{sec:bench_plants}

In this section, we analyze the proposed GFO method for tuning PMR controllers considering the same three plants used in \cite{art:lorenzini:2019:GFO_PR}, each one from a different class of plants defined in Subsection~\ref{subsec:class_plants}. They are represented by the following transfer functions:
\begin{equation*}
\label{eq:G_a_b_c}
G_{a}(s) {} = {} \frac{e^{-s}}{\left(s + 1\right)^2},\;\;G_b(s) {} = {} \frac{1}{\left(s + 1\right)^2}, \;\; G_c(s) {} = {} \frac{1}{s + 1}.
\end{equation*}

The first step in our procedure is to identify a frequency point for each plant, for which we performed the RAP experiment. This experiment yielded the parameters summarized in \tableref{tab:RAP_G}\footnote{All controller and plant parameters, as well as the performance measures, are given in the International System of Units.}. From these results and by using the tuning tables proposed in Subsection~\ref{sec:tables}, we designed PMR controllers for $N=1$, $3$, $5$ for the two sets of resonant frequencies and considering $\xi_n = 0$ to achieve asymptotic reference tracking. Then, we evaluated the closed-loop response to a sinusoidal reference (for the particular case $N=1$), and for periodic references formed by the first $N=3,~5$ modes of the sawtooth -- case (i) -- and square -- case (ii) -- wave signals. Thus, in these scenarios, $N$ represents both the harmonics number of the reference signal and the resonant modes in the controller under analysis. 

To assess the closed-loop response, performance criteria were evaluated in terms of the settling time $t_s$ -- considering a $2\%$ error tolerance --, in number of periods of the reference signal, that is, $n_s = \omega_r t_s/(2 \pi)$, and the maximum overshoot $M_o$ obtained through
\begin{equation*}
M_o = \max \left\{\frac{y_{max} - r_{max}}{r_{max}}, 0 \right\}\times 100 \%,
\end{equation*}
where $y_{max} = \max \left|y\right(t\left)\right|$ and $r_{max} = \max \left|r\right(t\left)\right|$.

Next, for each plant, we consider the PMR controllers in the scenarios: $N=1,~3,~5$ for both cases (i) and (ii). In each of these scenarios, we evaluate the GFO method for $\max(n\omega_r)=0.1\ov$ and $\max(n\omega_r)=0.9\ov$.

%%%%%%%%%%%%%%%%%%%%%%%%%%%%%%%%%%%%%%%%%%%%%%%%%%%%%%%%%%%%%%%%%%%%%%%%%%%%%%%%%%%%%%%%%%%%%%%%%%%%%%%%%%%%%%%%%%%%%%%%%%%%%%%%%%%%
%%%%%%%%%%%%%%%%%%%%%%%%%%%%%%%%%%%%%%%%%%%%%%%%%%%%%%%%%%%%%%%%%%%%%%%%%%%%%%%%%%%%%%%%%%%%%%%%%%%%%%%%%%%%%%%%%%%%%%%%%%%%%%%%%%%%

\subsection{Class A plant}\label{sec:plants_a}

\begin{figure}[t!]
	\centering
	\subfigure[$N = 1$, $\omega_r = 0.1 \omega_{u}$]
	{\includegraphics[width=0.802\columnwidth]{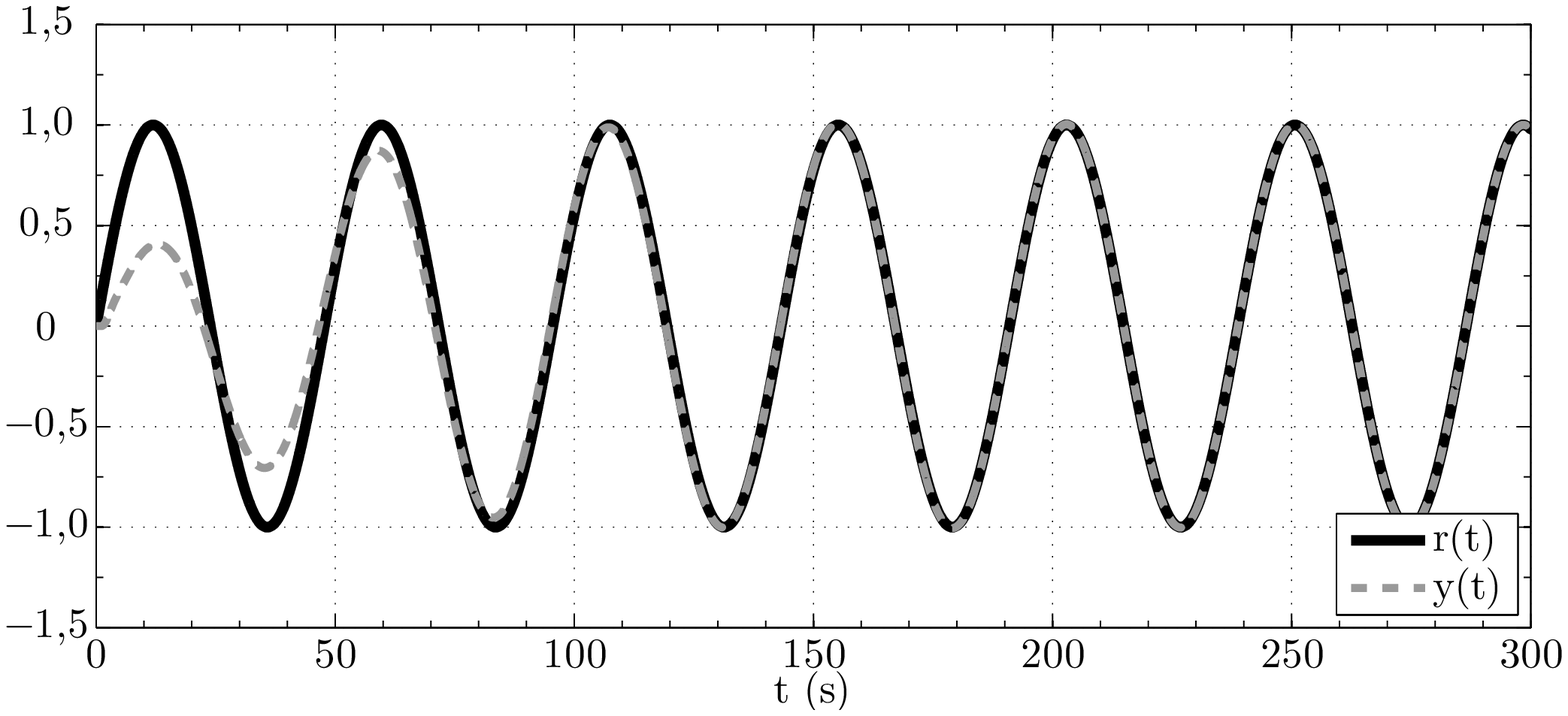}
		\label{fig:out_PRL_Ga_w_0_1}}
	\vskip-0.15cm
	\subfigure[$N = 1$, $\omega_r = 0.9 \omega_{u}$]
	{\includegraphics[width=0.802\columnwidth]{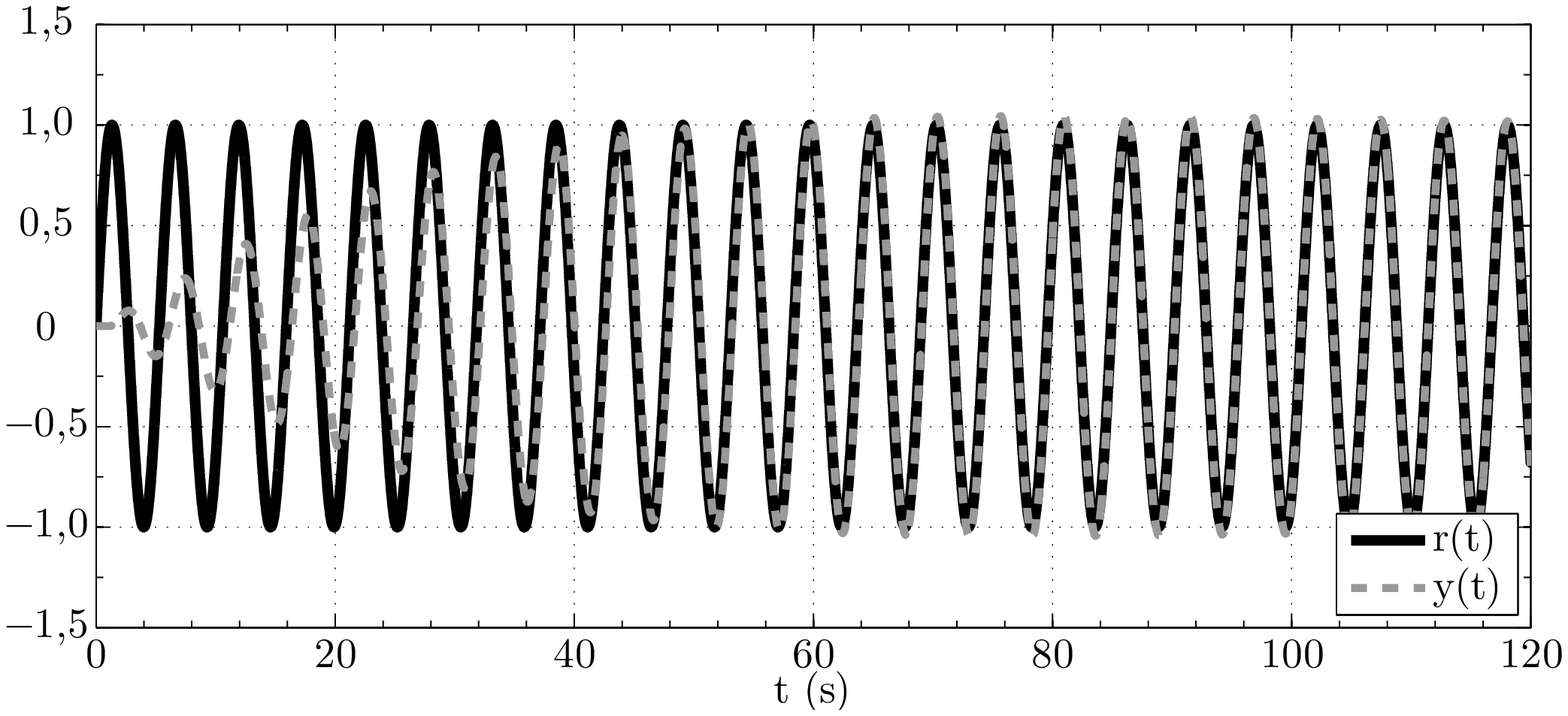}
		\label{fig:out_PRL_Ga_w_0_9}}
	\vskip-0.15cm
	\subfigure[$N = 5$, case (i), $\omega_r = 0.0200 \omega_{u}$]
	{\includegraphics[width=0.802\columnwidth]{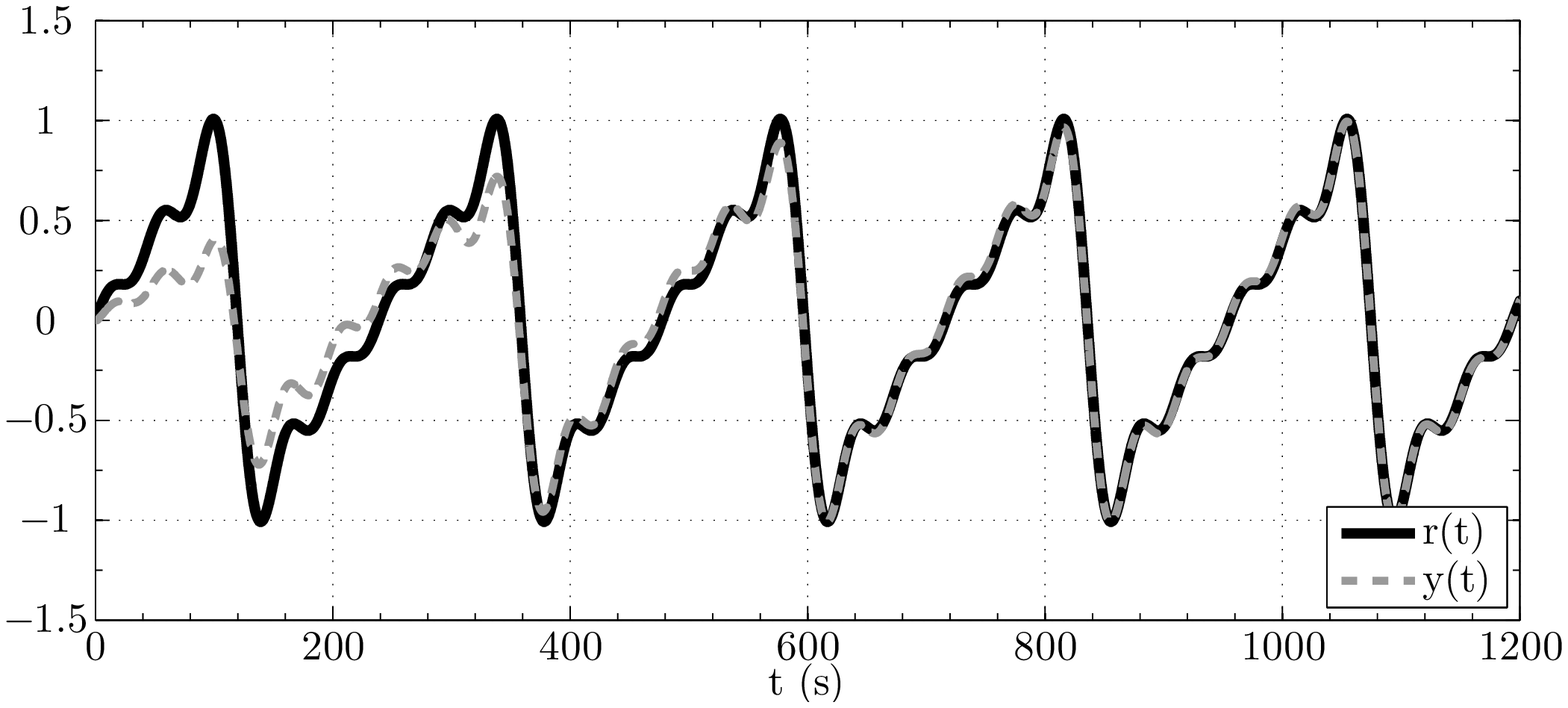}
		\label{fig:out_PMRL_5m_st_Ga_w_0_1}}
	\vskip-0.15cm
	\subfigure[$N = 5$, case (i), $\omega_r = 0.180 \omega_{u}$]
	{\includegraphics[width=0.802\columnwidth]{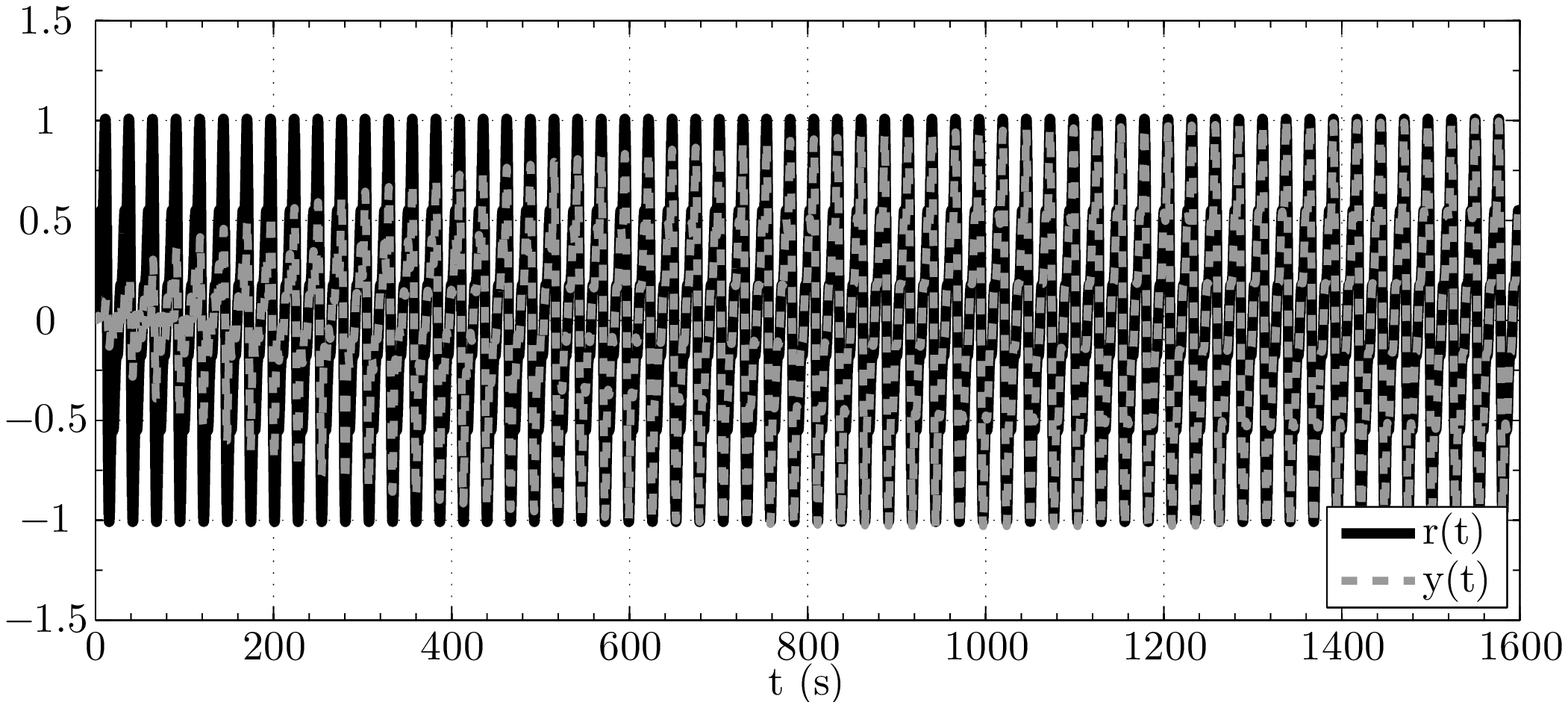}
		\label{fig:out_PMRL_5m_st_Ga_w_0_9}}
	\vskip-0.15cm
	\subfigure[$N = 5$, case (ii), $\omega_r = 0.0111 \omega_{u}$]
	{\includegraphics[width=0.802\columnwidth]{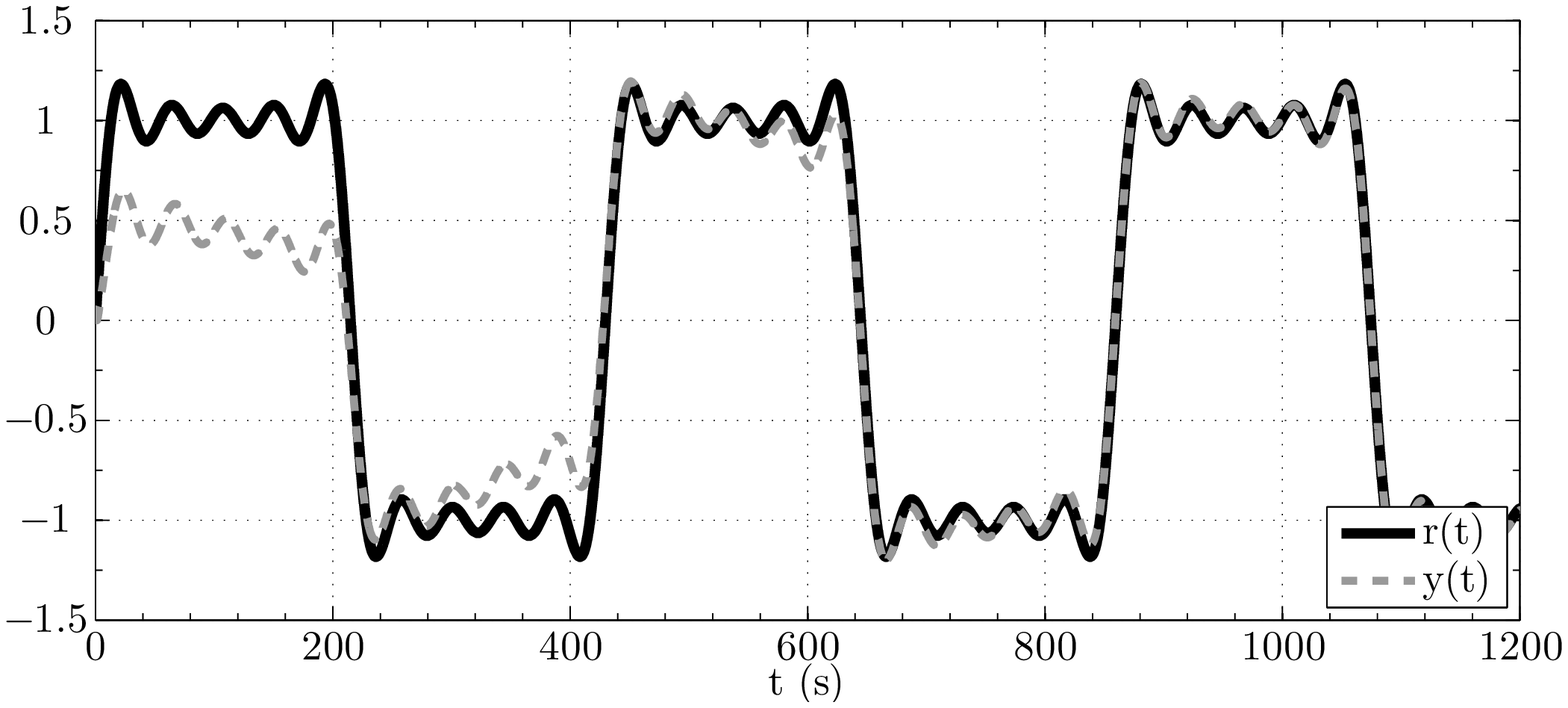}
		\label{fig:out_PMRL_5m_sq_Ga_w_0_1}}
	\vskip-0.15cm
	\subfigure[$N = 5$, case (ii), $\omega_r = 0.100 \omega_{u}$]
	{\includegraphics[width=0.802\columnwidth]{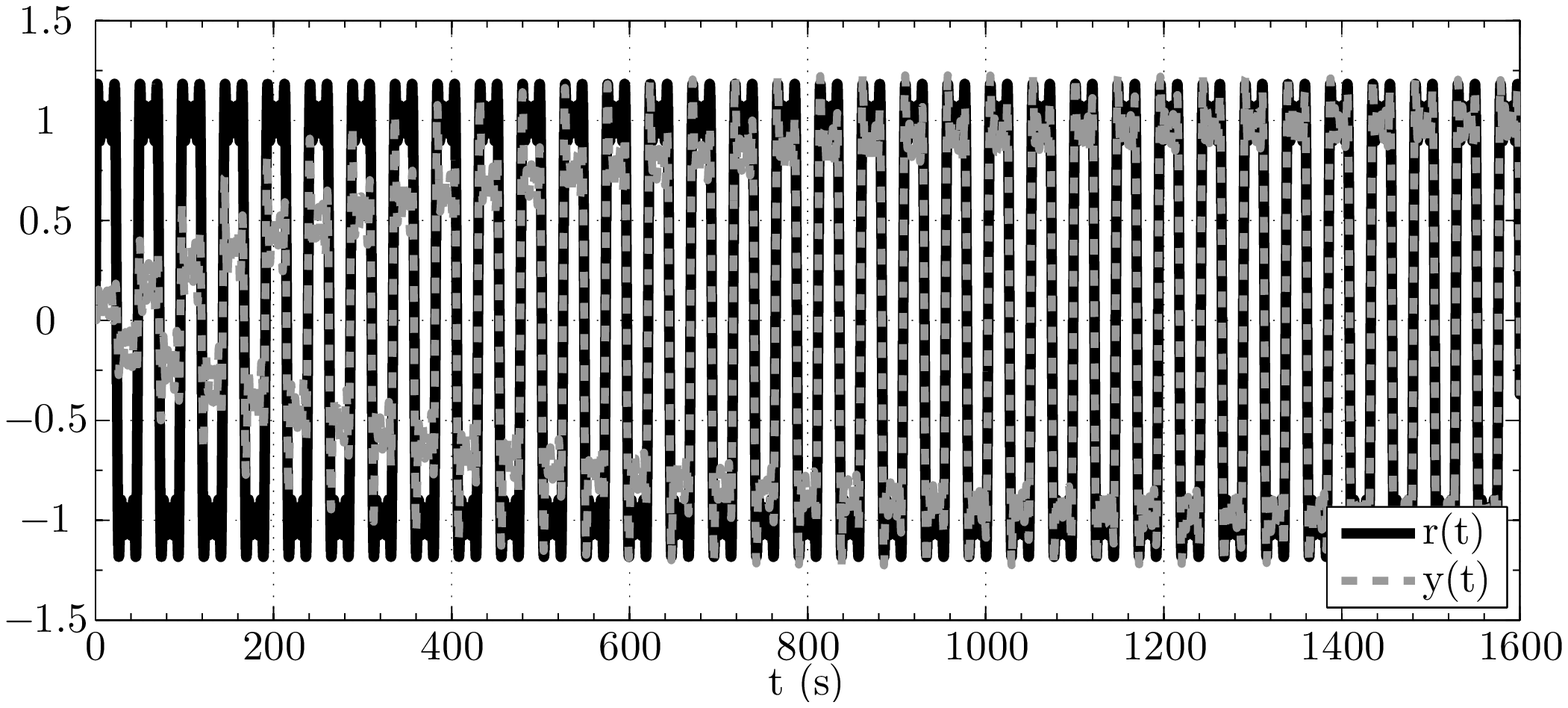}
		\label{fig:out_PMRL_5m_sq_Ga_w_0_9}}
	%\vskip-0.5cm
	\caption{Closed-loop response with the PMR controller and $G_a(s)$.} % (a) $N = 1$, $\omega_r = 0.1 \omega_{u}$. (b) $N = 1$, $\omega_r = 0.9 \omega_{u}$. (c) $N = 5$, case (i), $\omega_r = 0.0200 \omega_{u}$. (d) $N = 5$, case (i), $\omega_r = 0.180 \omega_{u}$. (e) $N = 5$, case (ii), $\omega_r = 0.0111 \omega_{u}$. (f) $N = 5$, case (ii), $\omega_r = 0.100 \omega_{u}$.}
	\label{fig:out_PRL_Ga}
\end{figure}

Initially, we consider the Class A plant $G_a(s)$. Based on the information obtained from the RAP experiment in Table \ref{tab:RAP_G}, application of the proposed tuning tables yielded the sets of controller parameters and performance measures summarized in \tableref{tab:result_Ga1} for the cases (i) and (ii). We notice that as the number of harmonics in the reference signal (and thus in the controller) increases, the settling time also increases. In contrast, the maximum overshoot is approximately the same for the same ratio $\omega_r/\omega_u$. But most importantly, we notice that all of these values are less than the imposed constraint of $15\%$. A set of closed-loop responses for each set of controller parameters and set of frequencies is shown in \figref{fig:out_PRL_Ga}.

\begin{table}[t!]
	\centering \renewcommand{\tabcolsep}{3pt} 
	\caption{Parameters for the RAP experiment} \label{tab:RAP_G}
	\small
	\begin{tabular}{c|ccccc|cc} \hline
		\textbf{Plant}      & \boldmath{$\gamma$}     & \boldmath{$\nu$}         & \boldmath{$d$}  & \boldmath{$A_\nu$} & \boldmath{$|F(j\omega_{\nu})|$} & \boldmath{$M_{\nu}$} & \boldmath{$\omega_{\nu}$}  \\ \hline  			
		$G_{a}(s)$ & $0\degree$    & $-180\degree$ & $1.3$ & $0.648$  & $1$                & $0.392$ & $1.32$       \\ %\hline 
		$G_{b}(s)$ & $-60\degree$  & $-120\degree$ & $2.4$ & $0.589$  & $0.757$            & $0.255$ & $1.69$       \\ %\hline
		$G_{c}(s)$ & $-120\degree$ & $-60\degree$  & $1.6$ & $0.532$  & $0.501$            & $0.501$ & $1.68$       \\ \hline 
	\end{tabular}
\end{table}

\begin{table}[t!]
	\centering \renewcommand{\tabcolsep}{0.8pt}
	\caption{Tuning and performance for $G_a(s)$} 	\label{tab:result_Ga1}
	\small
	\begin{tabular}{ccccccc} 	\hline
		\multirow{2}{*}{\textbf{Var.}}   & \multicolumn{6}{c}{\textbf{N}} \\ \cline{2-7}
		& \multicolumn{2}{c}{\textbf{1}}   & \multicolumn{2}{c}{\textbf{3}} & \multicolumn{2}{c}{\textbf{5}} \\ \hline 
		$k_a$                 & $2.50$    & $2.50$   & $2.50$      & $2.50$    & $2.50$      & $2.50$    \\ 
		$z_a$                 & $0.526$   & $0.526$  & $0.526$     & $0.526$   & $0.526$     & $0.526$   \\ 
		$p_a$                 & $3.29$    & $3.29$   & $3.29$      & $3.29$    & $3.29$      & $3.29$    \\ \hline 
		$\frac{\omega_r}{\omega_{u}}$ & $0.100$   & $0.900$  & $0.0333$    & $0.300$   & $0.0200$    & $0.180$   \\ 
		%$\xi_n$               & $0$       & $0$      & $0$         & $0$       & $0$         & $0$       \\
		$K_{p_1}$             & $1.01$    & $0.272$  & $1.02$      & $0.968$   & $1.02$      & $1.01$    \\ 
		$K_{r_{11}}$          & $0.162$   & $0.0311$ & $0.117$     & $0.107$   & $0.0702$    & $0.0680$  \\ 
		$K_{r_{21}}$          & $-0.0112$ & $-0.244$ & $-0.000997$ & $-0.0769$ & $-0.000134$ & $-0.0108$ \\ 
		$K_{p_2}$             &  --       &  --      & $0.999$     & $0.903$   & $1.00$      & $0.972$   \\ 
		$K_{r_{12}}$          &  --       &  --      & $0.0229$    & $0.0147$  & $0.0230$    & $0.0200$  \\ 
		$K_{r_{22}}$          &  --       &  --      & $-0.00146$  & $-0.107$  & $-0.000526$ & $-0.0415$ \\ 
		$K_{p_3}$             & --        & --       & $0.998$     & $0.552$   & $0.999$     & $0.927$   \\ 
		$K_{r_{13}}$          & --        & --       & $0.0228$    & $0.00437$ & $0.0229$    & $0.0163$  \\ 
		$K_{r_{23}}$          & --        & --       & $-0.00328$  & $-0.147$  & $-0.00118$  & $-0.0890$ \\ 
		$K_{p_4}$             & --        & --       & --          & --        & $0.999$     & $0.830$   \\ 
		$K_{r_{14}}$          & --        & --       & --          & --        & $0.0229$    & $0.0111$  \\ 
		$K_{r_{24}}$          & --        & --       & --          & --        & $-0.00210$  & $-0.142$  \\ 
		$K_{p_5}$             & --        & --       & --          & --        & $0.998$     & $0.552$   \\ 
		$K_{r_{15}}$          & --        & --       & --          & --        & $0.0228$    & $0.00437$ \\ 
		$K_{r_{25}}$          & --        & --       & --          & --        & $-0.00328$  & $-0.147$  \\  %\hline 
		$t_s$                 & $126$     & $115$    & $455$       & $321$     & $1008$      & $1457$    \\ 
		$n_s$                 & $2.6$     & $22$     & $3.2$       & $20$      & $4.2$       & $55$      \\ 
		$M_o$                 & $0.5$     & $4.4$    & $0.0$       & $1.3$     & $0.13$      & $2.2$     \\ \hline 
%	\end{tabular} 
%\end{table}	
%
%\begin{table}[t!]
%	\centering \renewcommand{\tabcolsep}{0.8pt}
%	\caption{Tuning and performance for $G_a(s)$ considering Case (ii)}
%	\label{tab:result_Ga2}
%	\small
%	\begin{tabular}{ccccccc} 	\hline
%		\multirow{2}{*}{\textbf{Var.}} & \multicolumn{6}{c}{\textbf{N}} \\ \cline{2-7}
%		& \multicolumn{2}{c}{\textbf{1}} & \multicolumn{2}{c}{\textbf{3}} & \multicolumn{2}{c}{\textbf{5}} \\ \hline 
		$\frac{\omega_r}{\omega_{u}}$ & $0.100$   & $0.900$  & $0.0200$    & $0.180$   & $0.0111$     & $0.100$    \\ 
%		$\xi_n$               & $0$       & $0$      & $0$         & $0$       & $0$          & $0$        \\
		$K_{p_1}$             & $1.01$    & $0.272$  & $1.02$      & $1.00$    & $1.02$       & $1.02$     \\ 
		$K_{r_{11}}$          & $0.162$   & $0.0311$ & $0.117$     & $0.113$   & $0.0702$     & $0.0695$   \\ 
		$K_{r_{21}}$          & $-0.0112$ & $-0.244$ & $-0.000359$ & $-0.0286$ & $-0.0000414$ & $-0.00335$ \\ 
		$K_{p_3}$             &  --       &  --      & $0.999$     & $0.927$   & $1.00$       & $0.982$    \\ %
		$K_{r_{13}}$          &  --       &  --      & $0.0229$    & $0.0163$  & $0.0230$     & $0.0210$   \\ 
		$K_{r_{23}}$          &  --       &  --      & $-0.00118$  & $-0.0890$ & $-0.000365$  & $-0.0291$  \\ 
		$K_{p_5}$             & --        & --       & $0.998$     & $0.555$   & $0.999$      & $0.940$    \\ 
		$K_{r_{15}}$          & --        & --       & $0.0228$    & $0.00437$ & $0.0230$     & $0.0173$   \\ 
		$K_{r_{25}}$          & --        & --       & $-0.00328$  & $-0.148$  & $-0.00199$   & $-0.136$   \\ 
		$K_{p_7}$             & --        & --       & --          & --        & $0.999$      & $0.846$    \\ 
		$K_{r_{17}}$          & --        & --       & --          & --        & $0.0229$     & $0.0117$   \\ 
		$K_{r_{27}}$          & --        & --       & --          & --        & $-0.00199$   & $-0.136$   \\ 
		$K_{p_9}$             & --        & --       & --          & --        & $0.998$      & $0.552$    \\ 
		$K_{r_{19}}$          & --        & --       & --          & --        & $0.0228$     & $0.00437$  \\ 
		$K_{r_{29}}$          & --        & --       & --          & --        & $-0.00328$   & $-0.147$   \\
%		$k_a$                 & $2.50$    & $2.50$   & $2.50$      & $2.50$    & $2.50$       & $2.50$     \\ 
%		$z_a$                 & $0.526$   & $0.526$  & $0.526$     & $0.526$   & $0.526$      & $0.526$    \\ 
%		$p_a$                 & $3.29$    & $3.29$   & $3.29$      & $3.29$    & $3.29$       & $3.29$     \\ \hline 
		$t_s$                 & $126$     & $115$    & $464$       & $428$     & $957$        & $1566$     \\ 
		$n_s$                 & $2.6$     & $22$     & $1.9$       & $16$      & $2.2$        & $33$       \\ 
		$M_o$                 & $0.5$     & $4.4$    & $0.0$       & $5.7$     & $1.3$        & $3.7$      \\ \hline 
	\end{tabular}  
\end{table}	

\begin{figure}[t!]
	\centering
	\subfigure[N = 1]
	{\includegraphics[width=0.75\columnwidth]{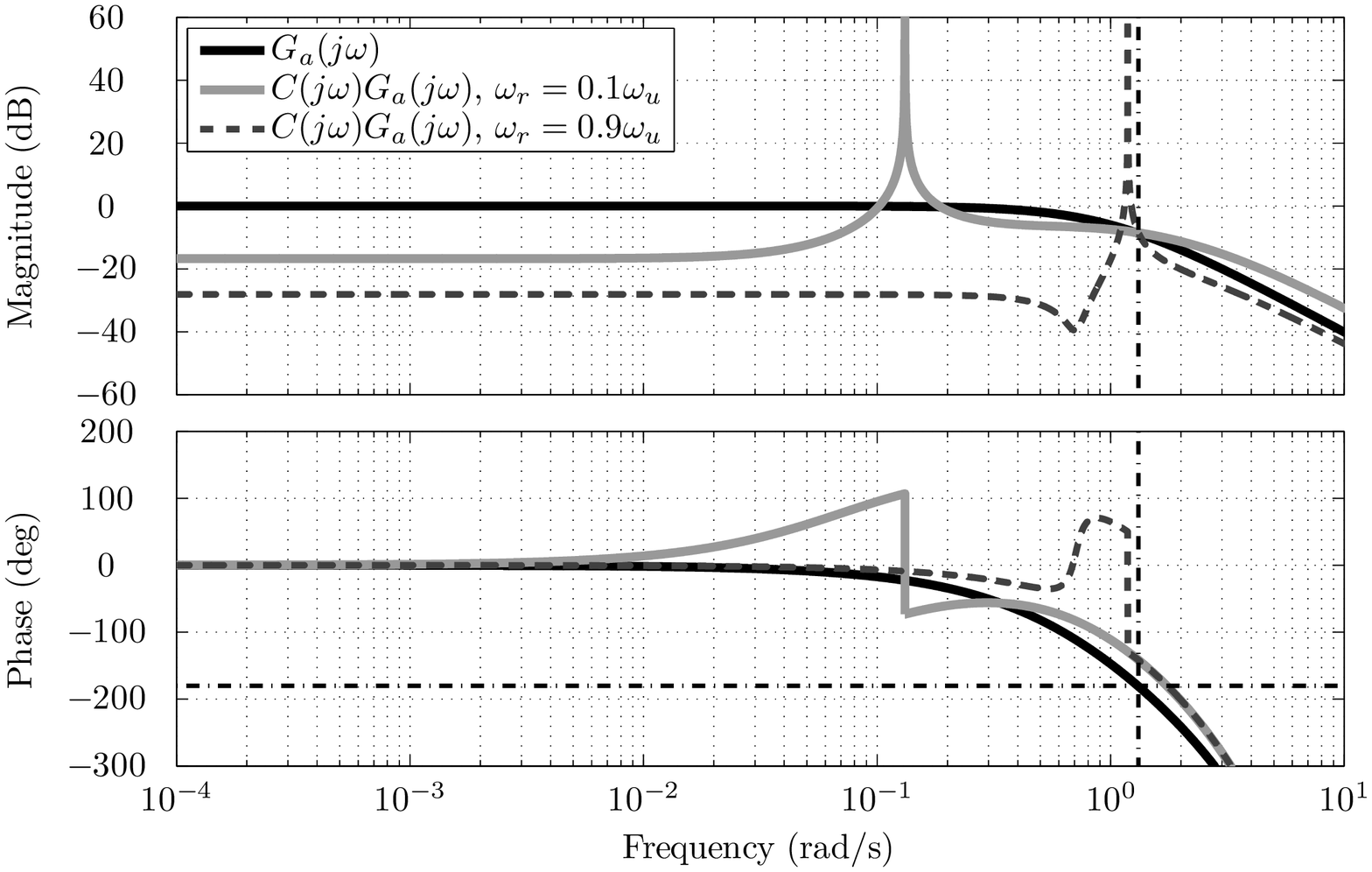}
		\label{fig:bode_Ga_prl_1m_wr_0_1_0_9}}
	\vskip-0.15cm
	\subfigure[$N = 5$, case (i)]
	{\includegraphics[width=0.75\columnwidth]{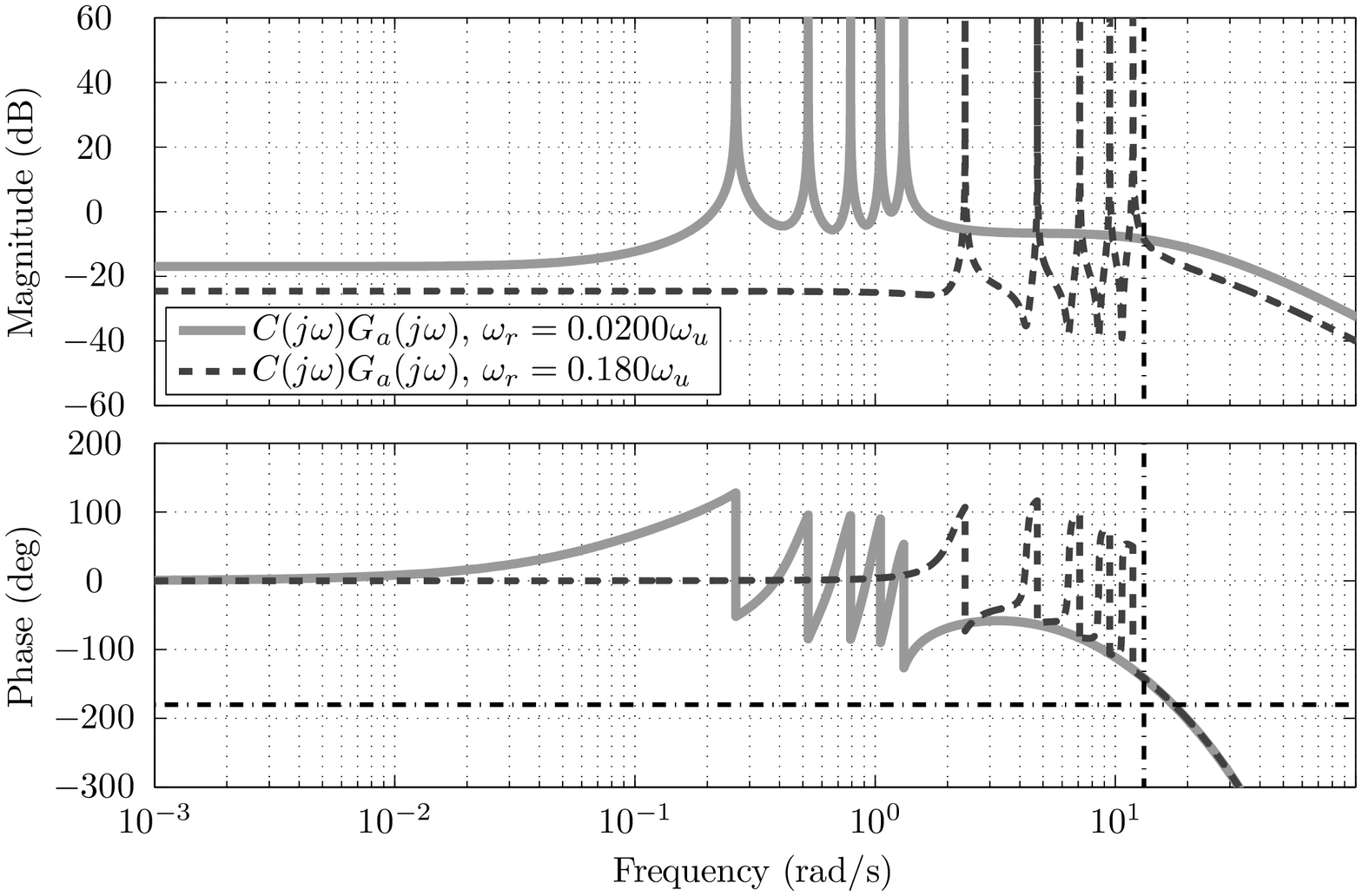}
		\label{fig:bode_Ga_pmrl_5m_wr_2wr_3wr_4wr_5wr_0_1_0_9}}
	\vskip-0.15cm
	\subfigure[$N = 5$, case (ii)]
	{\includegraphics[width=0.75\columnwidth]{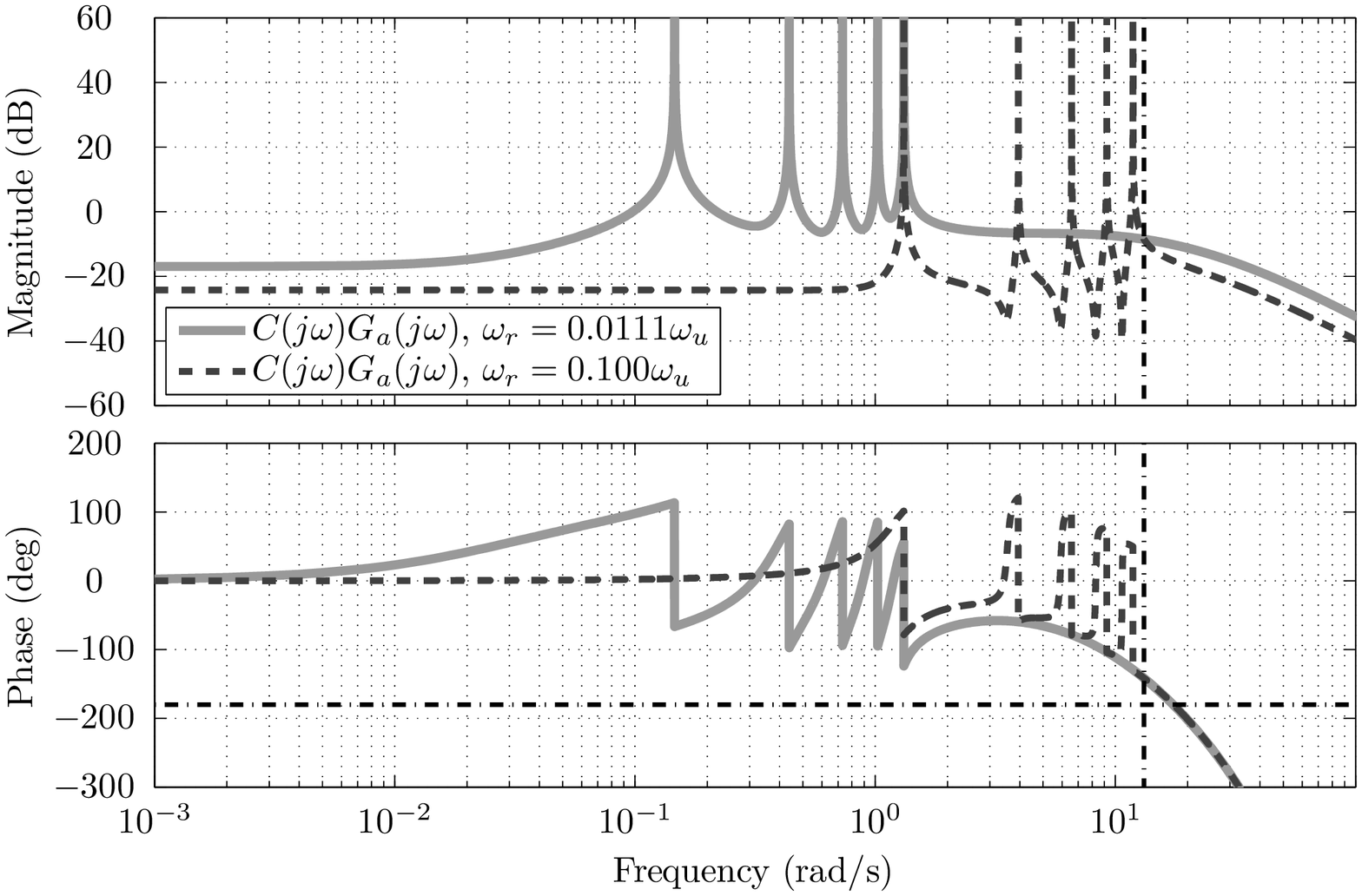}
		\label{fig:bode_Ga_pmrl_5m_wr_3wr_5wr_7wr_9wr_0_1_0_9}}
	%\vskip-0.15cm
	\caption{Frequency response of $G_a(j\omega)$ and $G_a(j\omega)C(j\omega)$. Black dashed lines are at $\omega_{u}=1.32$ and at $-180\degree$.}% (a) $N = 1$. (b) $N = 5$, case (i). (c) $N = 5$, case (ii).}
	\label{fig:bode_Ga_pmrl}
\end{figure}

A frequency response analysis helps to evaluate the proposed tuning. \figref{fig:bode_Ga_pmrl} shows the frequency response of the plant $G_a(j\omega)$ and the loop $C(j\omega)G_a(j\omega)$ for $N = 1,~5$ for the two sets of resonant modes and considering the reference frequencies $\max(n\omega_r)=0.1\omega_u$ and $\max(n\omega_r)=0.9\omega_u$. From these graphs, we can observe that the six controllers resulted in appropriate stability margins, even for the limit situation where the controllers have a resonance peak at $\max(n\omega_r)=0.9\omega_u$ and present very large gains in the range around the plant's ultimate frequency. 

\begin{figure}[t!]
	\centering
	%\subfigure[N = 1]{
		\includegraphics[width=0.65\columnwidth]{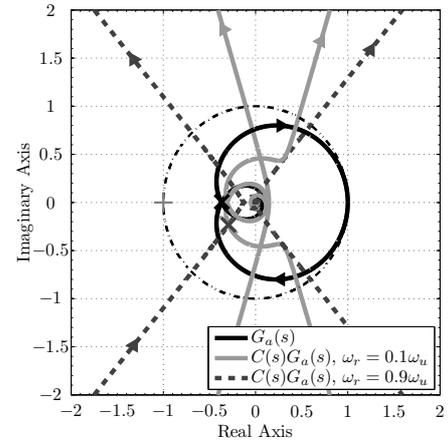}
		\label{fig:nyquist_Ga_pmrl_1m}
	%}
	%\vskip-0.15cm
	%\subfigure[$N = 5$, case (i)]{\includegraphics[width=0.76\columnwidth]{./Figures/5m/nyquist_Ga_pmra_5m_wr_2wr_3wr_4wr_5wr_0_1_0_9}
	%	\label{fig:nyquist_Ga_pmrl_5m_I}}
	%\vskip-0.15cm
	%\subfigure[$N = 5$, case (ii)]{\includegraphics[width=0.76\columnwidth]{./Figures/5m/nyquist_Ga_pmra_5m_wr_3wr_5wr_7wr_9wr_0_1_0_9}
	%	\label{fig:nyquist_Ga_pmrl_5m_II}}
	% D:\Dropbox\Doutorado\Ressonante\1_m\Gb\Gb_bode\ art_plot_bode_pr.m
	\caption{Nyquist diagrams of $G_a(s)$ and $G_a(s)C(s)$, which has $2$ turns with radius tending to infinity in the right-half of the complex plane.} %(a) $N = 1$. (b) $N = 5$, case (i). (c) $N = 5$, case (ii).}
	\label{fig:nyquist_Ga_pmrl}
\end{figure}
%
%The Nyquist diagrams of $G_a(s)$ and the loop transfer function $C(s)G_a(s)$ for each of these sets of resonance frequencies 
%are presented in \figref{fig:nyquist_Ga_pmrl}, where the point of frequency $\omega_u$ is marked with an ``X''. It should be noted that $C(s)$ has $2\times N$ poles at the imaginary axis in the frequencies $n\omega_r$, which implies that the Nyquist diagram of $C(s)G_a(s)$ has $2\times N$ turns with radius tending to infinity in the right-half complex plane. Besides, in these figures, it can be seen that the Nyquist diagrams of $C(s)G_a(s)$ do not encircle the point $-1+j0$ in any scenario, resulting in stable closed-loops.%

The Nyquist diagrams of $G_a(s)$ and the loop transfer function $C(s)G_a(s)$ for $N=1$ %each of these sets of resonance frequencies 
are presented in \figref{fig:nyquist_Ga_pmrl}, where the point of frequency $\omega_u$ is marked with an ``X''. It should be noted that, in this configuration with  $\xi_n = 0 $, $C(s)$ has $2\times N$ poles at the imaginary axis in the frequencies $n \omega_r$, which implies that the Nyquist diagram of $C(s)G_a(s)$ has $2\times N$ turns with radius tending to infinity in the right-half complex plane. Besides, in this figure, it can be seen that the Nyquist diagram of $C(s)G_a(s)$ do not encircle the point $-1+j0$, resulting in a stable closed-loop.% since $C(s)G_a(s)$ does not have poles in the right-half plane. 

The option for the PMR controller series configuration with a phase-lead block (for the Class A plants), in addition to the design variables -- the location $p$ in the complex plane and its decomposition in intermediate ones, and the relative restriction on the controller zeros -- guaranteed shifting of the plant ultimate point to $p=0.4e^{-j140.6^{\circ}}$, as designed in \eqref{eq:p_classA}. Furthermore, this control topology and design variables ensured adequate displacement of the plant ultimate point and its nearby points away from $-1+j0$, which guaranteed adequate stability margins and yielded suitable closed-loop performance with the six PMR controllers.

%%%%%%%%%%%%%%%%%%%%%%%%%%%%%%%%%%%%%%%%%%%%%%%%%%%%%%%%%%%%%%%%%%%%%%%%%%%%%%%%%%%%%%%%%%%%%%%%%%%%%%%%%%%%%%%%%%%%%%%%%%%%%%%%%%%%
%%%%%%%%%%%%%%%%%%%%%%%%%%%%%%%%%%%%%%%%%%%%%%%%%%%%%%%%%%%%%%%%%%%%%%%%%%%%%%%%%%%%%%%%%%%%%%%%%%%%%%%%%%%%%%%%%%%%%%%%%%%%%%%%%%%%

\subsection{Class B plant}\label{sec:plants_b}

\begin{table}[t!]
	\centering \renewcommand{\tabcolsep}{1.1pt}
	\caption{Tuning and performance  for $G_b(s)$}
	\label{tab:result_Gb1}
	\small
	\begin{tabular}{ccccccc} 	\hline
		\multirow{2}{*}{\textbf{Var.}} & \multicolumn{6}{c}{\textbf{N}} \\ \cline{2-7}
		& \multicolumn{2}{c}{\textbf{1}} & \multicolumn{2}{c}{\textbf{3}} & \multicolumn{2}{c}{\textbf{5}} \\ \hline  
		$\frac{\omega_r}{\omega_{120}}$ & $0.100$  & $0.900$ & $0.0333$   & $0.300$   & $0.0200$    & $0.180$   \\ 
		%$\xi_n$                 & $0$      & $0$     & $0$        & $0$       & $0$         & $0$       \\
		$K_{p_1}$               & $3.85$   & $1.22$  & $3.89$     & $3.81$    & $3.91$      & $3.88$    \\ 
		$K_{r_{11}}$            & $1.14$   & $0.220$ & $0.923$    & $0.841$   & $0.697$     & $0.675$   \\ 
		$K_{r_{21}}$            & $-0.0562$& $-1.44$ & $-0.00235$ & $-0.187$  & $-0.000850$ & $-0.0685$ \\ 
		$K_{p_2}$               & --       & --      & $0.999$    & $0.903$   & $1.00$      & $0.972$   \\ 
		$K_{r_{12}}$            & --       & --      & $0.0295$   & $0.0190$  & $0.0296$    & $0.0258$  \\ 
		$K_{r_{22}}$            & --       & --      & $-0.00242$ & $-0.177$  & $-0.000871$ & $-0.0686$ \\ 
		$K_{p_3}$               & --       & --      & $0.998$    & $0.552$   & $0.999$     & $0.927$   \\ 
		$K_{r_{13}}$            & --       & --      & $0.0293$   & $0.00563$ & $0.0295$    & $0.0210$  \\ 
		$K_{r_{23}}$            & --       & --      & $-0.00544$ & $-0.244$  & $-0.00196$  & $-0.147$  \\ 
		$K_{p_4}$               & --       & --      & --         & --        & $0.999$     & $0.830$   \\ 
		$K_{r_{14}}$            & --       & --      & --         & --        & $0.0294$    & $0.0143$  \\ 
		$K_{r_{24}}$            & --       & --      & --         & --        & $-0.00348$  & $-0.234$  \\ 
		$K_{p_5}$               & --       & --      & --         & --        & $0.998$     & $0.552$   \\ 
		$K_{r_{15}}$            & --       & --      & --         & --        & $0.0293$    & $0.00563$ \\ 
		$K_{r_{25}}$            & --       & --      & --         & --        & $-0.00544$  & $-0.244$  \\ 
		$t_s$                   & $27$     & $27$    & $96$       & $46$      & $150$       & $117$     \\ 
		$n_s$                   & $0.74$   & $6.5$   & $0.87$     & $3.7$     & $0.81$      & $5.7$     \\ 
		$M_o$                   & $1.9$    & $1.6$   & $0.070$    & $0$       & $0.39$      & $0.046$   \\ \hline 
%	\end{tabular} 
%\end{table}
%
%\begin{table}[t!]
%	\centering \renewcommand{\tabcolsep}{1.1pt}
%	\caption{Tuning and performance  for $G_b(s)$ and Case (ii)}
%	\label{tab:result_Gb2}
%	\small
%	\begin{tabular}{ccccccc} 	\hline
%		\multirow{2}{*}{\textbf{Var.}} & \multicolumn{6}{c}{\textbf{N}} \\ \cline{2-7}
%		& \multicolumn{2}{c}{\textbf{1}} & \multicolumn{2}{c}{\textbf{3}} & \multicolumn{2}{c}{\textbf{5}} \\ \hline 
		$\frac{\omega_r}{\omega_{120}}$ & $0.100$  & $0.900$ & $0.0200$    & $0.180$   & $0.0111$    & $0.100$   \\ 
%		$\xi_n$                 & $0$      & $0$     & $0$         & $0$       & $0$         & $0$       \\
		$K_{p_1}$               & $3.85$   & $1.22$  & $3.89$      & $3.86$    & $3.91$      & $3.90$    \\ 
		$K_{r_{11}}$            & $1.14$   & $0.220$ & $0.923$     & $0.894$   & $0.698$     & $0.691$   \\ 
		$K_{r_{21}}$            & $-0.0562$& $-1.44$ & $-0.000846$ & $-0.0681$ & $-0.000263$ & $-0.0212$ \\ 
		$K_{p_3}$               & --       & --      & $0.999$     & $0.927$   & $1.00$      & $0.982$   \\ 
		$K_{r_{13}}$            & --       & --      & $0.0295$    & $0.0210$  & $0.0296$    & $0.0270$  \\ 
		$K_{r_{23}}$            & --       & --      & $-0.00196$  & $-0.147$  & $-0.000605$ & $-0.0481$ \\ 
		$K_{p_5}$               & --       & --      & $0.998$     & $0.552$   & $0.999$     & $0.940$   \\ 
		$K_{r_{15}}$            & --       & --      & $0.0293$    & $0.00563$ & $0.0295$    & $0.0222$  \\ 
		$K_{r_{25}}$            & --       & --      & $-0.00544$  & $-0.244$  & $-0.00329$  & $-0.226$  \\ 
		$K_{p_7}$               & --       & --      & --          & --        & $0.999$     & $0.846$   \\ 
		$K_{r_{17}}$            & --       & --      & --          & --        & $0.0294$    & $0.0151$  \\ 
		$K_{r_{27}}$            & --       & --      & --          & --        & $-0.00329$  & $-0.226$  \\ 
		$K_{p_9}$               & --       & --      & --          & --        & $0.998$     & $0.552$   \\ 
		$K_{r_{19}}$            & --       & --      & --          & --        & $0.0293$    & $0.00563$ \\ 
		$K_{r_{29}}$            & --       & --      & --          & --        & $-0.00544$  & $-0.244$  \\ 
		$t_s$                   & $27$     & $27$    & $218$       & $50$      & $519$       & $129$     \\ 
		$n_s$                   & $0.74$   & $6.5$   & $1.2$       & $2.4$     & $1.6$       & $3.5$     \\ 
		$M_o$                   & $1.9$    & $1.6$   & $3.3$       & $0.022$   & $4.2$       & $0.32$    \\ \hline 
	\end{tabular} 
\end{table}	

Consider now the Class B plant $G_b(s)$ and the information obtained from the RAP experiment for this plant shown in Table \ref{tab:RAP_G}. Using the proposed tuning tables, we achieved the sets of controller parameters and performance measures presented in \tableref{tab:result_Gb1} for the cases (i) and (ii). Notice the same behavior, as in Class A, for the settling time, which increases with $N$, and for the maximum overshoot, which is less than the aimed value of $15\%$. %A set of closed-loop responses for each set of controller parameters and set of frequencies is shown in \figref{fig:out_PMR_Gb}. %. In contrast, the maximum overshoot values have a peak for $N=1$, and then it decreases for about half the value for other $N$, given the same ratio $\omega_r/\omega_u$ for case (i); on the other hand, for case (ii), the maximum overshoot is higher when $n\omega_r$ is close to $\omega_u$.  A set of closed-loop responses for each set of controller parameters and set of frequencies is shown in \figref{fig:out_PMR_Gb}.

\subsection{Class C plant}\label{sec:plants_c}

Finally, we analyze the GFO method applied to the Class C plant $G_c(s)$. For this plant, based on the information presented in Table \ref{tab:RAP_G} and application of the proposed tuning tables, we obtained the sets of controller parameters and performance measures listed in \tableref{tab:result_Gc1} for the cases (i) and (ii). We notice the same behavior as in classes A and B for the settling time, given the same $N$ and ratio $\omega_r/\omega_\nu$, and also for the maximum overshoot values, which are less than $15\%$, as desired. %A set of closed-loop responses for each set of controller parameters and set of frequencies is presented in \figref{fig:out_PMR_Gc}.

\begin{table}[t!]
	\centering \renewcommand{\tabcolsep}{1.2pt}
	\caption{Tuning and performance for $G_c(s)$}
	\label{tab:result_Gc1}
	\small
	\begin{tabular}{ccccccc} 	\hline
		\multirow{2}{*}{\textbf{Var.}} & \multicolumn{6}{c}{\textbf{N}} \\ \cline{2-7}
		& \multicolumn{2}{c}{\textbf{1}} & \multicolumn{2}{c}{\textbf{3}} & \multicolumn{2}{c}{\textbf{5}} \\ \hline 
		$\frac{\omega_r}{\omega_{60}}$ & $0.100$   & $0.900$  & $0.0333$   & $0.300$   & $0.0200$    & $0.180$   \\ 
		%$\xi_n$                & $0$       & $0$      & $0$        & $0$       & $0$         & $0$       \\
		$K_{p_1}$              & $1.71$    & $0.332$  & $1.76$     & $1.73$    & $1.80$      & $1.78$    \\ 
		$K_{r_{11}}$           & $1.66$    & $0.319$  & $1.57$     & $1.43$    & $1.47$      & $1.42$    \\ 
		$K_{r_{21}}$           & $-0.0479$ & $-0.751$ & $-0.00105$ & $-0.0832$ & $-0.000384$ & $-0.0309$ \\ 
		$K_{p_2}$              & --        & --       & $0.999$    & $0.903$   & $1.00$      & $0.972$   \\ 
		$K_{r_{12}}$           & --        & --       & $0.0292$   & $0.0188$  & $0.0293$    & $0.0255$  \\ 
		$K_{r_{22}}$           & --        & --       & $-0.00237$ & $-0.173$  & $-0.000853$ & $-0.0672$ \\ 
		$K_{p_3}$              & --        & --       & $0.998$    & $0.552$   & $0.999$     & $0.927$   \\ 
		$K_{r_{13}}$           & --        & --       & $0.0290$   & $0.00557$ & $0.0292$    & $0.0208$  \\ 
		$K_{r_{23}}$           & --        & --       & $-0.00532$ & $-0.239$  & $-0.00192$  & $-0.144$  \\ 
		$K_{p_4}$              & --        & --       & --         & --        & $0.999$     & $0.830$   \\ 
		$K_{r_{14}}$           & --        & --       & --         & --        & $0.0291$    & $0.0141$  \\ 
		$K_{r_{24}}$           & --        & --       & --         & --        & $-0.00341$  & $-0.230$  \\ 
		$K_{p_5}$              & --        & --       & --         & --        & $0.998$     & $0.552$   \\ 
		$K_{r_{15}}$           & --        & --       & --         & --        & $0.0290$    & $0.00557$ \\ 
		$K_{r_{25}}$           & --        & --       & --         & --        & $-0.00532$  & $-0.239$  \\ 
		$t_s$                  & $94$      & $24$     & $101$      & $49$      & $168$       & $113$     \\ 
		$n_s$                  & $2.5$     & $5.8$    & $0.90$     & $4.0$     & $0.9$       & $5.4$     \\ 
		$M_o$                  & $6.3$     & $0$      & $1.2$      & $0.018$    & $1.3$       & $0$       \\ \hline 
%	\end{tabular} 
%\end{table}	
%
%\begin{table}[t!]
%	\centering \renewcommand{\tabcolsep}{1.2pt}
%	\caption{Tuning and performance  for $G_c(s)$ and Case (ii)}
%	\label{tab:result_Gc2}
%	\small
%	\begin{tabular}{ccccccc} 	\hline
%		\multirow{2}{*}{\textbf{Var.}} & \multicolumn{6}{c}{\textbf{N}} \\ \cline{2-7}
%		& \multicolumn{2}{c}{\textbf{1}} & \multicolumn{2}{c}{\textbf{3}} & \multicolumn{2}{c}{\textbf{5}} \\ \hline  
		$\frac{\omega_r}{\omega_{60}}$  & $0.100$   & $0.900$  & $0.0200$    & $0.180$   & $0.0111$    & $0.100$    \\ 
%		$\xi_n$                & $0$       & $0$      & $0$         & $0$       & $0$         & $0$        \\
		$K_{p_1}$              & $1.71$    & $0.332$  & $1.76$      & $1.75$    & $1.80$      & $1.79$     \\ 
		$K_{r_{11}}$           & $1.66$    & $0.319$  & $1.57$      & $1.52$    & $1.47$      & $1.45$     \\ 
		$K_{r_{21}}$           & $-0.0479$ & $-0.751$ & $-0.000377$ & $-0.0303$ & $-0.000118$ & $-0.00957$ \\ 
		$K_{p_3}$              & --        & --       & $0.999$     & $0.927$   & $1.00$      & $0.982$    \\ 
		$K_{r_{13}}$           & --        & --       & $0.0292$    & $0.0208$  & $0.0293$    & $0.0267$   \\ 
		$K_{r_{23}}$           & --        & --       & $-0.00192$  & $-0.144$  & $-0.000592$ & $-0.0471$  \\ 
		$K_{p_5}$              & --        & --       & $0.998$     & $0.552$   & $0.999$     & $0.940$    \\ 
		$K_{r_{15}}$           & --        & --       & $0.0290$    & $0.00557$ & $0.0292$    & $0.0220$   \\ 
		$K_{r_{25}}$           & --        & --       & $-0.00532$  & $-0.239$  & $-0.00322$  & $-0.221$   \\ 
		$K_{p_7}$              & --        & --       & --          & --        & $0.999$     & $0.846$    \\ 
		$K_{r_{17}}$           & --        & --       & --          & --        & $0.0291$    & $0.0150$   \\ 
		$K_{r_{27}}$           & --        & --       & --          & --        & $-0.00322$  & $-0.221$   \\ 
		$K_{p_9}$              & --        & --       & --          & --        & $0.998$     & $0.552$    \\ 
		$K_{r_{19}}$           & --        & --       & --          & --        & $0.0290$    & $0.00557$  \\ 
		$K_{r_{29}}$           & --        & --       & --          & --        & $-0.00532$  & $-0.239$   \\ 
		$t_s$                  & $94$      & $24$     & $340$       & $55$      & $736$       & $116$      \\ 
		$n_s$                  & $2.5$     & $5.8$    & $1.8$       & $2.6$     & $2.2$       & $3.1$      \\ 
		$M_o$                  & $6.3$     & $0$      & $3.0$       & $0.054$   & $2.9$       & $0.014$    \\ \hline 
	\end{tabular} 
\end{table}

\section{Conclusions} \label{sec:conc}

In this paper we proposed an innovative development to the GFO method for tuning PMR controllers. This method is based on the identification of the most appropriate point of the frequency response for each class of plants through the RAP experiment. We developed four sets of tuning formulas to obtain appropriate stability margins and closed-loop performance for each class of plants: three sets for tuning the first order harmonic for each class, and one set for tuning the other higher order harmonics for all classes. We also introduced a phase-lead block for plants that possess ultimate point in order to improve the phase margin.
The proposed methodology was validated considering a wide variety of plants, periodic references with different compositions of multiple integer frequencies below the plant's identified frequency and also PMR controllers with up to five resonant modes tuned at these frequencies. Good closed-loop performance (in terms of settling time and maximum overshoot) and robustness (which is obtained through appropriate stability margins) have been achieved for all such cases. We highlight this is an easily implementable and easy computable model-free methodology for the multifrequency resonant controllers design that requires only a simple RAP experiment on the process.

%Thus, We presented an easily understandable and easily computable, model-free methodology that is applicable to multifrequency resonant controllers. %Further developments of the GFO method to consider proportional-integral-multi-resonant controllers are relevant topics for future research.

\section*{References}

\bibliography{bibliography}

\begin{thebibliography}{20}
\expandafter\ifx\csname natexlab\endcsname\relax\def\natexlab#1{#1}\fi
\providecommand{\url}[1]{\texttt{#1}}
\providecommand{\href}[2]{#2}
\providecommand{\path}[1]{#1}
\providecommand{\DOIprefix}{doi:}
\providecommand{\ArXivprefix}{arXiv:}
\providecommand{\URLprefix}{URL: }
\providecommand{\Pubmedprefix}{pmid:}
\providecommand{\doi}[1]{\href{http://dx.doi.org/#1}{\path{#1}}}
\providecommand{\Pubmed}[1]{\href{pmid:#1}{\path{#1}}}
\providecommand{\bibinfo}[2]{#2}
\ifx\xfnm\undefined \def\xfnm[#1]{\unskip,\space#1}\fi
%Type = Article
\bibitem[{{\AA}str{\"o}m and H{\"a}gglund(1984)}]{art:astrom:1984:rele}
\bibinfo{author}{{\AA}str{\"o}m\xfnm[ K.J.]},
  \bibinfo{author}{H{\"a}gglund\xfnm[ T.]}.
\newblock \bibinfo{title}{Automatic tuning of simple regulators with
  specifications on phase and amplitude margins}.
\newblock \bibinfo{journal}{Automatica}
  \bibinfo{year}{1984};\bibinfo{volume}{20}(\bibinfo{number}{5}):\bibinfo{pages}{645--651}.
%Type = Book
\bibitem[{{\AA}str{\"o}m and H{\"a}gglund(1995)}]{book:pid:astrom:1995:pid}
\bibinfo{author}{{\AA}str{\"o}m\xfnm[ K.J.]},
  \bibinfo{author}{H{\"a}gglund\xfnm[ T.]}.
\newblock \bibinfo{title}{PID controllers: theory, design, and tuning}.
\newblock \bibinfo{address}{Research Triangle Park, NC, USA}:
  \bibinfo{publisher}{ISA}, \bibinfo{year}{1995}.
%Type = Article
\bibitem[{Bazanella et~al.(2017)Bazanella, Pereira and
  Parraga}]{art:bazanella:2017:PID-rele-foi}
\bibinfo{author}{Bazanella\xfnm[ A.S.]}, \bibinfo{author}{Pereira\xfnm[
  L.F.A.]}, \bibinfo{author}{Parraga\xfnm[ A.]}.
\newblock \bibinfo{title}{A new method for \protect{PID} tuning including
  plants without ultimate frequency}.
\newblock \bibinfo{journal}{IEEE Transactions on Control Systems Technology}
  \bibinfo{year}{2017};\bibinfo{volume}{25}(\bibinfo{number}{2}):\bibinfo{pages}{637--644}.
%Type = Article
\bibitem[{{Castilla} et~al.(2009){Castilla}, {Miret}, {Matas}, {Garcia de
  Vicuna} and {Guerrero}}]{art:Castilla:2009:PMR_xi}
\bibinfo{author}{{Castilla}\xfnm[ M.]}, \bibinfo{author}{{Miret}\xfnm[ J.]},
  \bibinfo{author}{{Matas}\xfnm[ J.]}, \bibinfo{author}{{Garcia de
  Vicuna}\xfnm[ L.]}, \bibinfo{author}{{Guerrero}\xfnm[ J.M.]}.
\newblock \bibinfo{title}{Control design guidelines for single-phase
  grid-connected photovoltaic inverters with damped resonant harmonic
  compensators}.
\newblock \bibinfo{journal}{IEEE Transactions on Industrial Electronics}
  \bibinfo{year}{2009};\bibinfo{volume}{56}(\bibinfo{number}{11}):\bibinfo{pages}{4492--4501}.
%Type = Article
\bibitem[{Francis and Wonham(1975)}]{art:contr:francis:1975}
\bibinfo{author}{Francis\xfnm[ B.]}, \bibinfo{author}{Wonham\xfnm[ W.]}.
\newblock \bibinfo{title}{The internal model principle for linear multivariable
  regulators}.
\newblock \bibinfo{journal}{Applied Mathematics and Optimization}
  \bibinfo{year}{1975};\bibinfo{volume}{2}(\bibinfo{number}{2}):\bibinfo{pages}{170--194}.
%Type = Article
\bibitem[{Habibullah et~al.(2017)Habibullah, Pota and
  Petersen}]{art:res:Habibullah:2017:vibracao}
\bibinfo{author}{Habibullah\xfnm[ H.]}, \bibinfo{author}{Pota\xfnm[ H.R.]},
  \bibinfo{author}{Petersen\xfnm[ I.R.]}.
\newblock \bibinfo{title}{A novel control approach for high precision
  positioning of a piezoelectric tube scanner}.
\newblock \bibinfo{journal}{IEEE Transactions on Automation Science and
  Engineering}
  \bibinfo{year}{2017};\bibinfo{volume}{14}(\bibinfo{number}{1}):\bibinfo{pages}{325--336}.
%Type = Article
\bibitem[{{Hans} et~al.(2020){Hans}, {Schumacher}, {Chou} and
  {Wang}}]{art:Hans:2020:Design_PMR}
\bibinfo{author}{{Hans}\xfnm[ F.]}, \bibinfo{author}{{Schumacher}\xfnm[ W.]},
  \bibinfo{author}{{Chou}\xfnm[ S.]}, \bibinfo{author}{{Wang}\xfnm[ X.]}.
\newblock \bibinfo{title}{Design of multifrequency proportional-resonant
  current controllers for voltage-source converters}.
\newblock \bibinfo{journal}{IEEE Transactions on Power Electronics}
  \bibinfo{year}{2020};:\bibinfo{pages}{1--1}.
%Type = Article
\bibitem[{{Lascu} et~al.(2007){Lascu}, {Asiminoaei}, {Boldea} and
  {Blaabjerg}}]{art:Lascu:2007:Active:Power:Filters}
\bibinfo{author}{{Lascu}\xfnm[ C.]}, \bibinfo{author}{{Asiminoaei}\xfnm[ L.]},
  \bibinfo{author}{{Boldea}\xfnm[ I.]}, \bibinfo{author}{{Blaabjerg}\xfnm[
  F.]}.
\newblock \bibinfo{title}{High performance current controller for selective
  harmonic compensation in active power filters}.
\newblock \bibinfo{journal}{IEEE Transactions on Power Electronics}
  \bibinfo{year}{2007};\bibinfo{volume}{22}(\bibinfo{number}{5}):\bibinfo{pages}{1826--1835}.
%Type = Article
\bibitem[{Lorenzini et~al.(2019)Lorenzini, Bazanella, Pereira and
  Gon\-\c{c}al\-ves~da Silva}]{art:lorenzini:2019:GFO_PID}
\bibinfo{author}{Lorenzini\xfnm[ C.]}, \bibinfo{author}{Bazanella\xfnm[ A.S.]},
  \bibinfo{author}{Pereira\xfnm[ L.F.A.]},
  \bibinfo{author}{Gon\-\c{c}al\-ves~da Silva\xfnm[ G.R.]}.
\newblock \bibinfo{title}{The generalized forced oscillation method for tuning
  {PID} controllers}.
\newblock \bibinfo{journal}{ISA Transactions}
  \bibinfo{year}{2019};\bibinfo{volume}{87}:\bibinfo{pages}{68--87}.
%Type = Article
\bibitem[{{Lorenzini} et~al.(2020){Lorenzini}, {Pereira} and
  {Bazanella}}]{art:lorenzini:2019:GFO_PR}
\bibinfo{author}{{Lorenzini}\xfnm[ C.]}, \bibinfo{author}{{Pereira}\xfnm[
  L.F.A.]}, \bibinfo{author}{{Bazanella}\xfnm[ A.S.]}.
\newblock \bibinfo{title}{A generalized forced oscillation method for tuning
  proportional-resonant controllers}.
\newblock \bibinfo{journal}{IEEE Transactions on Control Systems Technology}
  \bibinfo{year}{2020};\bibinfo{volume}{28}(\bibinfo{number}{3}):\bibinfo{pages}{1108--1115}.
%Type = Article
\bibitem[{Moheimani and Vautier(2005)}]{art:res:Moheimani:2005:ress_struc}
\bibinfo{author}{Moheimani\xfnm[ S.O.R.]}, \bibinfo{author}{Vautier\xfnm[
  B.J.G.]}.
\newblock \bibinfo{title}{Resonant control of structural vibration using
  charge-driven piezoelectric actuators}.
\newblock \bibinfo{journal}{IEEE Transactions on Control Systems Technology}
  \bibinfo{year}{2005};\bibinfo{volume}{13}(\bibinfo{number}{6}):\bibinfo{pages}{1021--1035}.
%Type = Article
\bibitem[{Pereira and Bazanella(2015)}]{art:pereira:2015:PR-ZN}
\bibinfo{author}{Pereira\xfnm[ L.F.A.]}, \bibinfo{author}{Bazanella\xfnm[
  A.S.]}.
\newblock \bibinfo{title}{Tuning rules for proportional resonant controllers}.
\newblock \bibinfo{journal}{IEEE Transactions on Control Systems Technology}
  \bibinfo{year}{2015};\bibinfo{volume}{23}(\bibinfo{number}{5}):\bibinfo{pages}{2010--2017}.
%Type = Article
\bibitem[{Pereira et~al.(2014)Pereira, Flores, Bonan, Coutinho and {Gomes da
  Silva Jr.}}]{art:res:pereira:2014:mr}
\bibinfo{author}{Pereira\xfnm[ L.F.A.]}, \bibinfo{author}{Flores\xfnm[ J.V.]},
  \bibinfo{author}{Bonan\xfnm[ G.]}, \bibinfo{author}{Coutinho\xfnm[ D.F.]},
  \bibinfo{author}{{Gomes da Silva Jr.}\xfnm[ J.M.]}.
\newblock \bibinfo{title}{Multiple resonant controllers for uninterruptible
  power supplies - a systematic robust control design approach}.
\newblock \bibinfo{journal}{IEEE Transactions on Industrial Electronics}
  \bibinfo{year}{2014};\bibinfo{volume}{61}(\bibinfo{number}{3}):\bibinfo{pages}{1528--1538}.
%Type = Article
\bibitem[{{Tao} et~al.(2020){Tao}, {Zhu}, {Xu}, {Li} and
  {Zhu}}]{art:Tao:2020:Nano:PMR}
\bibinfo{author}{{Tao}\xfnm[ Y.]}, \bibinfo{author}{{Zhu}\xfnm[ Z.]},
  \bibinfo{author}{{Xu}\xfnm[ Q.]}, \bibinfo{author}{{Li}\xfnm[ H.]},
  \bibinfo{author}{{Zhu}\xfnm[ L.]}.
\newblock \bibinfo{title}{Tracking control of nanopositioning stages using
  parallel resonant controllers for high-speed nonraster sequential scanning}.
\newblock \bibinfo{journal}{IEEE Transactions on Automation Science and
  Engineering} \bibinfo{year}{2020};:\bibinfo{pages}{1--11}.
%Type = Article
\bibitem[{Teodorescu et~al.(2006)Teodorescu, Blaabjerg, Liserre and
  Loh}]{art:res:Teodorescu:2006:ups}
\bibinfo{author}{Teodorescu\xfnm[ R.]}, \bibinfo{author}{Blaabjerg\xfnm[ F.]},
  \bibinfo{author}{Liserre\xfnm[ M.]}, \bibinfo{author}{Loh\xfnm[ P.C.]}.
\newblock \bibinfo{title}{Proportional-resonant controllers and filters for
  grid-connected voltage-source converters}.
\newblock \bibinfo{journal}{IEE Proceedings - Electric Power Applications}
  \bibinfo{year}{2006};\bibinfo{volume}{153}(\bibinfo{number}{5}):\bibinfo{pages}{750--762}.
%Type = Book
\bibitem[{Tepljakov(2017)}]{book:tepljakov:2017:fractional}
\bibinfo{author}{Tepljakov\xfnm[ A.]}.
\newblock \bibinfo{title}{Fractional-order modeling and control of dynamic
  systems}.
\newblock \bibinfo{address}{New York}: \bibinfo{publisher}{Springer Berlin
  Heidelberg}, \bibinfo{year}{2017}.
%Type = Article
\bibitem[{{Trinh} and {Lee}(2013)}]{art:Trinh:2013:APF_Res}
\bibinfo{author}{{Trinh}\xfnm[ Q.]}, \bibinfo{author}{{Lee}\xfnm[ H.]}.
\newblock \bibinfo{title}{An advanced current control strategy for three-phase
  shunt active power filters}.
\newblock \bibinfo{journal}{IEEE Transactions on Industrial Electronics}
  \bibinfo{year}{2013};\bibinfo{volume}{60}(\bibinfo{number}{12}):\bibinfo{pages}{5400--5410}.
%Type = Article
\bibitem[{{Xin} et~al.(2018){Xin}, {Mattavelli}, {Yao}, {Yang}, {Blaabjerg} and
  {Loh}}]{art:Xin:2018}
\bibinfo{author}{{Xin}\xfnm[ Z.]}, \bibinfo{author}{{Mattavelli}\xfnm[ P.]},
  \bibinfo{author}{{Yao}\xfnm[ W.]}, \bibinfo{author}{{Yang}\xfnm[ Y.]},
  \bibinfo{author}{{Blaabjerg}\xfnm[ F.]}, \bibinfo{author}{{Loh}\xfnm[ P.C.]}.
\newblock \bibinfo{title}{Mitigation of grid-current distortion for
  {LCL}-filtered voltage-source inverter with inverter-current feedback
  control}.
\newblock \bibinfo{journal}{IEEE Transactions on Power Electronics}
  \bibinfo{year}{2018};\bibinfo{volume}{33}(\bibinfo{number}{7}):\bibinfo{pages}{6248--6261}.
%Type = Article
\bibitem[{{Yepes} et~al.(2011){Yepes}, {Freijedo}, {Lopez} and
  {Doval-Gandoy}}]{art:Yepes:2011:Nyquist}
\bibinfo{author}{{Yepes}\xfnm[ A.G.]}, \bibinfo{author}{{Freijedo}\xfnm[
  F.D.]}, \bibinfo{author}{{Lopez}\xfnm[ O.]},
  \bibinfo{author}{{Doval-Gandoy}\xfnm[ J.]}.
\newblock \bibinfo{title}{Analysis and design of resonant current controllers
  for voltage-source converters by means of {Nyquist} diagrams and sensitivity
  function}.
\newblock \bibinfo{journal}{IEEE Transactions on Industrial Electronics}
  \bibinfo{year}{2011};\bibinfo{volume}{58}(\bibinfo{number}{11}):\bibinfo{pages}{5231--5250}.
%Type = Article
\bibitem[{Ziegler and Nichols(1942)}]{art:pid:ZN:1942}
\bibinfo{author}{Ziegler\xfnm[ J.G.]}, \bibinfo{author}{Nichols\xfnm[ N.B.]}.
\newblock \bibinfo{title}{Optimum settings for automatic controllers}.
\newblock \bibinfo{journal}{Transactions ASME}
  \bibinfo{year}{1942};\bibinfo{volume}{64}(\bibinfo{number}{11}):\bibinfo{pages}{759--768}.

\end{thebibliography}

\end{document}